\documentclass[journal]{IEEEtran}

\usepackage[colorlinks,urlcolor=blue,linkcolor=blue,citecolor=blue]{hyperref}

\usepackage{color,array}

\usepackage{graphicx}
\usepackage{mathtools}
\usepackage{amssymb}
\DeclareMathOperator*{\argmin}{argmin}
\usepackage{dsfont}

\newcommand{\R}{\mathbb{R}}
\usepackage{cite}
\usepackage{amsmath,amssymb,amsfonts}
\usepackage{amsmath}
\usepackage{mathrsfs}

\usepackage{graphicx}
\usepackage{textcomp}
\usepackage[table,xcdraw]{xcolor}
\usepackage{ragged2e}
\usepackage{bbm}
\usepackage{multirow}
\usepackage[linesnumbered, ruled]{algorithm2e}
\usepackage{algpseudocode}
\usepackage[caption=false, font=footnotesize]{subfig}
\usepackage{mathtools} 
\usepackage[hang,flushmargin]{footmisc}
\usepackage{lipsum}
\makeatletter
\newcommand{\algorithmfootnote}[2][\footnotesize]{%
  \let\old@algocf@finish\@algocf@finish% Store algorithm finish macro
  \def\@algocf@finish{\old@algocf@finish% Update finish macro to insert "footnote"
    \leavevmode\rlap{\begin{minipage}{\linewidth}
    #1#2
    \end{minipage}}%
  }%
}
\usepackage[flushleft]{threeparttable}
\usepackage{etoolbox}
\appto\TPTnoteSettings{\footnotesize}

\newcommand{\thickhline}{%
    \noalign {\ifnum 0=`}\fi \hrule height 1pt
    \futurelet \reserved@a \@xhline
}
\newcolumntype{"}{@{\hskip\tabcolsep\vrule width 1pt\hskip\tabcolsep}}

\usepackage{mathtools}
\DeclarePairedDelimiter\ceil{\lceil}{\rceil}
\DeclarePairedDelimiter\floor{\lfloor}{\rfloor}

\usepackage{tabu}

\makeatletter
\def\ps@IEEEtitlepagestyle{%
  \def\@oddfoot{\mycopyrightnotice}%
  \def\@evenfoot{}%
}
\def\mycopyrightnotice{%
  {\footnotesize “This work has been submitted to the IEEE for possible publication. Copyright may be transferred without notice, after which this version may no longer be accessible.”\hfill}% <--- Change here
  \gdef\mycopyrightnotice{}% just in case
}

\begin{document}

\title{CovSegNet: A Multi Encoder-Decoder Architecture for Improved Lesion Segmentation of COVID-19 Chest CT Scans}

\author{Tanvir Mahmud, \IEEEmembership{Student Member, IEEE}, Md Awsafur Rahman, \IEEEmembership{Student Member, IEEE}, Shaikh Anowarul Fattah, \IEEEmembership{Senior~Member, IEEE}, and~Sun-Yuan~Kung, \IEEEmembership{Life~Fellow, IEEE} 

\thanks{T.~Mahmud, M.~A.~Rahman, and S.~A.~Fattah are with the Department
of Electrical and Electronic Engineering, Bangladesh University of Engineering and Technology, Dhaka-1000, Bangladesh e-mail: (tanvirmahmud@eee.buet.ac.bd, mdawsafurrahman@ug.eee.buet.ac.bd, and fattah@eee.buet.ac.bd).}
\thanks{S-Y.~Kung is with the Department
of Electrical Engineering, Princeton University, USA e-mail: kung@princeton.edu}

\thanks{This paragraph will include the Associate Editor who handled your paper.}}

\markboth{Tanvir Mahmud \MakeLowercase{\textit{et al.}}: CovSegNet Architecture}
{}

\maketitle

\begin{abstract}
Automatic lung lesions segmentation of chest CT scans is considered a pivotal stage towards accurate diagnosis and severity measurement of COVID-19. 
Traditional U-shaped encoder-decoder architecture and its variants suffer from diminutions of contextual information in pooling/upsampling operations with increased semantic gaps among encoded and decoded feature maps as well as instigate vanishing gradient problems for its sequential gradient propagation
that result in sub-optimal performance.
Moreover, operating with 3D CT-volume poses further limitations due to the exponential increase of computational complexity making the optimization difficult.
In this paper, an automated COVID-19 lesion segmentation scheme is proposed utilizing a highly efficient neural network architecture, namely CovSegNet, to overcome these limitations. 
Additionally, a two-phase training scheme is introduced where a deeper 2D-network is employed for generating ROI-enhanced CT-volume followed by a shallower 3D-network for further enhancement with more contextual information without increasing computational burden.
Along with the traditional vertical expansion of Unet, we have introduced horizontal expansion with multi-stage encoder-decoder modules for achieving optimum performance. Additionally, multi-scale feature maps are integrated into the scale transition process to overcome the loss of contextual information. Moreover,  a multi-scale fusion module is introduced with a pyramid fusion scheme to reduce the semantic gaps between subsequent encoder/decoder modules while facilitating the parallel optimization for efficient gradient propagation.
Outstanding performances have been achieved in three publicly available datasets that largely outperform other state-of-the-art approaches. 
The proposed scheme can be easily extended for achieving optimum segmentation performances in a wide variety of applications.
\end{abstract}

\begin{IEEEImpStatement}
With lower sensitivity (60-70\%), elongated testing time, and a dire shortage of testing kits, traditional RT-PCR based COVID-19 diagnostic scheme heavily relies on post-CT based manual inspection for further investigation. Hence, automating the process of infected lesions extraction from chest-CT volumes will be major progress for faster accurate diagnosis of COVID-19. 
However, in challenging conditions with diffused, blurred, and varying shaped edges of COVID-19 lesions, conventional approaches fail to provide precise segmentation of lesions that can be deleterious for false estimation and loss of information.
The proposed scheme incorporating an efficient neural network architecture (CovSegNet) overcomes the limitations of traditional approaches that provide significant improvement of performance (8.4\% in averaged dice measurement scale) over two datasets. 
Therefore, this scheme can be an effective, economical tool for the physicians for faster infection analysis to greatly reduce the spread and massive death toll of this deadly virus through mass-screening.
\end{IEEEImpStatement}


\section{Introduction}
\IEEEPARstart{W}{ith} the recent outbreak of Coronavirus disease-2019 (COVID-19), the world has experienced an unprecedented number of deaths with a major collapse in the healthcare system throughout the world~\cite{c1,c3}. Early diagnosis is the primary concern to control this global pandemic at this stage for its extreme infectious nature~\cite{c2}. Though Reverse transcription-polymerase chain reaction (RT-PCR) is considered as the gold standard for diagnosing COVID-19, its longer time requirement, lower sensitivity with a massive shortage of test-kits have already engendered the extreme urgency of alternative automated diagnostic schemes~\cite{rt1,rt2}. Due to the wide applicability of the artificial intelligence (AI) tools in numerous clinical diagnostic measures, it has enormous potential to expedite the diagnostic process of COVID-19 through automated analysis and interpretation of the clinical record~\cite{a1,a2}.

Chest radiography has already been proven to be an effective source for COVID diagnostics due to its major implications relating to various levels of lung infections~\cite{ct1}. Computer tomography (CT) scan and chest X-ray have been extensively explored in the literature to establish an automated AI-based COVID diagnostic scheme~\cite{cov,x1,ct2}. Despite the easier access to chest X-ray, CT scans are more widely accepted due to its finer details leveraging the accurate diagnosis of COVID infections. Precise segmentation of lung lesions in chest  CT scans is one of the most demanding and challenging aspects for faster diagnosis of COVID-19 due to the shortage of annotated data, diverse levels of infections, and novel types and characteristics of the infections~\cite{inf}.  

Processing 3D CT volume at a whole increases computational complexity exponentially that makes the optimization and convergence more difficult limiting the architectural diversity of the network. The most widely used alternative of 3D-processing is to operate separately on 2D-slices extracted from the CT-volume~\cite{inf, u1, u2, u3, u4}. However, such slice-based processing loses inter-slice contextual information that results in sub-optimal performance. In~\cite{ua1,  ua2, vnet, cvnet}, smaller sub-volumes are extracted from the original 3D volumes to minimize the computational burden as well as to utilize 3D contextual information. However, such methods suffer from inter-volume contextual information loss by considering a smaller portion of the whole set at a time as well as increases complexity to process sub-volume level prediction into the final result.

A wide variety of approaches have been introduced in recent years for segmenting the region-of-interest in diverse applications. In~\cite{fcn}, a fully connected network (FCN) is introduced that produces multiple scales of encoded feature maps and reconstructs the segmentation mask utilizing these encoded representations. In~\cite{unet}, Unet architecture is introduced by integrating an inverted decoder module following the encoder module to gradually reconstruct the mask that gains much popularity over the years. However, several architectural limitations of Unet are identified that provides suboptimal performance.
\begin{itemize}
    \item  The skip connection introduced in Unet generates semantic gap between corresponding feature scale of encoder-decoder modules, which mainly arises from the direct concatenation of two semantically dissimilar feature maps. As the encoder module encodes the input image gradually into more generalized feature representation, it contains richer details compared to the corresponding decoded feature map which contains more information for the reconstruction of the final segmentation mask. These existing semantic gaps between corresponding encoder and decoder feature maps make the optimization process more difficult to converge for such direct concatenations through skip connections.
    %Moreover, increased semantic gaps are generated among multi-scale feature maps of the encoder-decoder modules due to its sequential nature without parallel inter-linking of multi-scale features.

    \item Contextual information loss occurs in traditional pooling/strided convolution-based downsampling operations that become more eminent with deeper architecture. Such downsampling operations are mainly carried out for generating more generalized, sparser feature representation with increased channels and reduced spatial resolution of the feature map. However, these operations also lead to loss of contextual information that greatly rises with the increase of vertical depth of the network. Similarly, the traditional upsampling operations fail to properly incorporate contextual information.
%    It relies on a deeper vertical structure with a single stage of encoder and decoder module. 

   \item The vanishing gradient problem rises in a deeper structure for sequential optimization of multi-scale features. This problem mainly arises from the difficulty of gradient propagation through the deep stack of convolutional layers. 
   %In Unet, skip interconnection sequentially connects corresponding encoder and decoder layer providing a path for the backpropagation of gradients. 
   Along with the incorporation of additional levels in the encoder and decoder stacks to make the network deeper, it becomes increasingly difficult to back-propagate the gradients through these levels for propagating through longer sequential paths that make the optimization of the deeper layers more difficult. Hence, this problem reduces the effective contributions of the deeper layers of the encoder and decoder modules for improper optimization. 
    
    \item Simplistic sequential convolutional layers are integrated into each level of encoder/decoder modules that lack enough architectural diversity to extract features from a broader spectrum, which is mainly caused by the linear propagation of gradients that reduces the impact of prior convolutional layers at each level for diminishing gradients. It lacks opportunity for the proper reuse of extracted features in the successive convolutions and lacks parallelism among convolutional layers required for better optimization, which lower the diversity of features generated at different levels of the network.  
    
\end{itemize}

Different architectural modifications have been explored in recent years to overcome some of these limitations. To increase the diversity of operations at each scale of feature maps, numerous established network building blocks are integrated in encoder/decoder module, e.g. residual block~\cite{res}, dense block~\cite{dunet}, inception block~\cite{inc}, dilated residual block~\cite{dense}, and multi-res block~\cite{mrunet}. To reduce the semantic gap between a particular scale of encoder and decoder, a residual path is proposed in MultiResUnet architecture instead of a direct skip connection of Unet~\cite{mrunet}. However, the semantic gap generated between multi-scale feature maps of encoder and decoder modules still persists. In Unet++~\cite{unet++}, a nested stack of convolutional layers is introduced to reduce the semantic gaps. But, it increases computational complexity considerably which makes convergence difficult. 
In~\cite{vnet}, Vnet is proposed that utilizes residual building blocks in Unet architecture, while  in~\cite{cvnet}, cascaded-Vnet is presented for performance improvement that utilizes a dual-stack of the cascaded encoder-decoder module. Nevertheless, with existing numerous architectural limitations of traditional U-shaped architecture in each stage, it increases semantic gaps with the additional encoding-decoding stage as well as increases vanishing gradient issues with contextual information loss that open up opportunities for further optimization.

\begin{figure*}[t]
    \centering
    \includegraphics[scale=0.18]{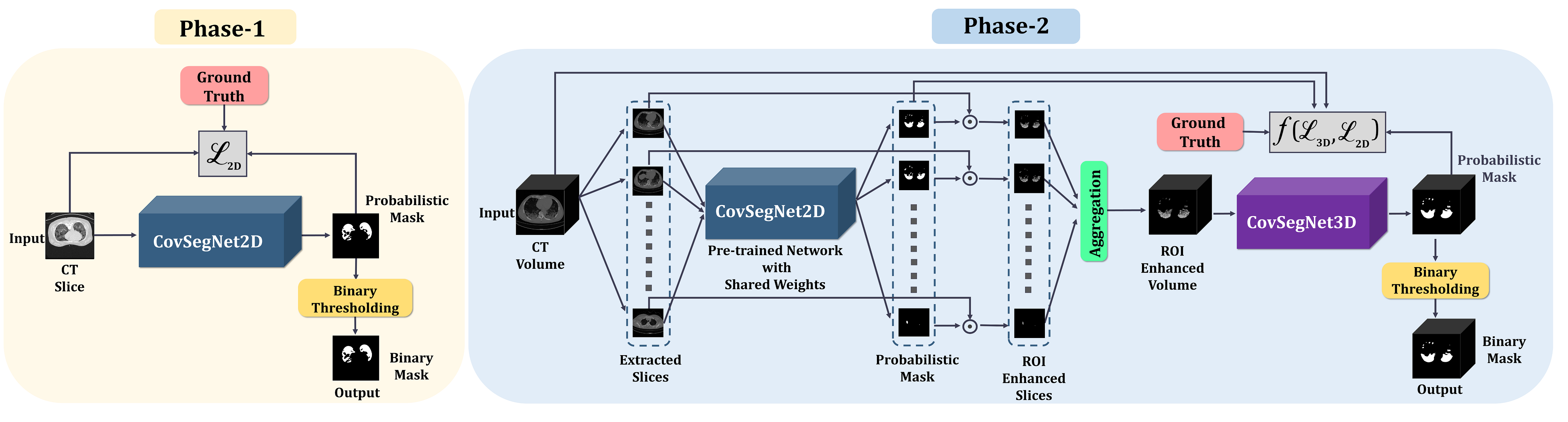}
    \caption{\textbf{Workflow of the proposed scheme for segmenting lung lesions of COVID-19 in CT volume. In phase-1, deeper CovSegNet2D is trained and optimized with CT-slices. In phase-2, further joint optimization is carried out where  pre-trained CovSegNet2D is fine-tuned for generating the ROI-enhanced CT volume while shallower form of CovSegNet3D is trained for more precise volumetric segmentation.}}
    \label{f1}
\end{figure*}

In this paper, an improved, automated scheme is proposed for precise lesion segmentation of COVID-19 chest CT volumes by overcoming the limitations of traditional approaches with a novel deep neural network architecture, named as CovSegNet. The major contributions of this work are summarized below:

\begin{enumerate}
    \item Along with the opportunity of vertical expansion, a horizontal expansion strategy is introduced in the CovSegNet architecture. In the vertical expansion mechanism, the encoder and decoder modules are deepened, while in horizontal expansion, several encoding-decoding stages are integrated. As discussed earlier, loss of contextual information occurs when the network is vertically expanded through subsequent downsampling operations, though it provides the opportunity for improved generalization through incorporating features from higher levels. Whereas, the horizontal expansion mechanism assists to integrate more detailed features at each level for finer reconstruction that helps to recover the loss of contextual information. As a result, it provides the opportunity to increase generalization while exploiting the available contextual information through an optimal combination of horizontal and vertical stages. 
    
    %Hence, several stages of sequential encoder-decoder modules can be integrated through a joint optimization scheme along with a deeper encoder-decoder structure for achieving optimum performance. 
    %This opportunity helps to reduce the information loss through pooling scheme with reduced vertical depth along with increased horizontal depth of the network.
    
    \item For further replenishing the loss of contextual information in traditional pooling/upsampling operations, a scale transition scheme is introduced in the encoder/decoder module by incorporating multi-scale feature maps from preceding levels. This scale transition scheme also improves the gradient flow across different feature scales of a particular encoder/decoder module.
    
    \item For reducing semantic gaps among corresponding feature scales of the encoder-decoder modules, a multi-scale fusion module is introduced in between successive encoder-decoder modules. This module fuses multi-scale feature representations, generated at preceding encoder/decoder modules through pyramid fusion scheme, to generate representational features with reduced semantic gap and improved contextual information for the following decoder/encoder module, instead of directly connecting corresponding feature scales like Unet. Moreover, this module establishes parallel linkage among multi-scale feature maps of subsequent encoder-decoder modules that greatly improve the gradient flow across the network and helps to reduce the vanishing gradient problem.
    
    %improving gradient flow by initializing the parallel optimization of numerous encoder-decoder modules. A multi-scale fusion module is introduced for establishing parallel linkage among multi-scale feature maps as well as for reducing the semantic gap among subsequent encoder/decoder modules, where multiscale processing is introduced utilizing a pyramid fusion scheme with diversified scaling operations.
    %by establishing a dense interconnection among multi-scale feature maps aggregating all preceding stages.
    %\item A pyramid fusion scheme is introduced to incorporate effective fusion on aggregated feature vector through employing multi-scale processing utilizing numerous upsampling/downsampling operations with diverse windows.
    
    \item A multi-phase training approach is introduced for integrating the advantages of both the 2D and 3D data processing scheme to reach the optimum performance. 2D processing provides faster processing with lower memory consumption while losing inter-slice contextual information. Whereas, 3D processing exploits both the intra-slice and inter-slice contextual information while increasing the computational burden. The proposed multi-phase training solves this problem by integrating a deeper variant of CovSegNet2D followed by a much shallower variant of CovSegNet3D for exploiting all possible contextual information while limiting the computational burden.
    %The ROI-enhancement operation is carried out utilizing a deeper variant of CovSegNet2D, pre-trained in the initial training phase, while further joint optimization is incorporated for 3D-data processing on ROI-enhanced volume with a shallower form of CovSegNet3D.

    \item The proposed CovSegNet architecture is designed in a modular and structured way that can be adapted to its lightweight, shallow form to reduce complicacy with considerable performance as well as can be made very deep to increase diversity for incorporating finer details. This generic design provides more flexibility for tuning the design parameters in a wide variety of applications.

    \item Extensive experimentations have been carried out to validate the effectiveness of the proposed scheme on two publicly available datasets containing chest CT scans from COVID-19 patients. Moreover, to validate the wide applicability of the proposed architecture, experimental results on a challenging, non-clinical, semantic segmentation dataset are also provided.

\end{enumerate}

%a deep neural network architecture is proposed, named as CovSegNet, for automated segmentations of the lung lesions in chest CT volumes by overcoming all the limitations presented in traditional architecture. 

%importance of covid segmentation

%challenges in 3d processing, problems of traditional unet, fcn networks and strategies

%dual decoder-encoder
%generalized architecture
%effecive 3d processing scheme
%multi-scale fusion scheme
%pyramid fusion scheme

%The proposed CovSegNet architecture is designed in a modular and structured way that can be adapted to its lightweight, shallow form to reduce complicacy while it can be made very deep for integating more optimization. As the 2-dimensional convolutional operations are more efficient in terms of computational complexity while available 2D-slices are considerably higher compared to 3D CT volumes, a deep variant of CovSegNet2D can be employed to operate on CT slices.  On the contrary, 3D convolutional operations are expensive in terms of computations along with fewer amount of available CT volumes limit the applicability of deep 3D-network in the process.

\begin{figure*}[t]
    \centering
    \includegraphics[scale=0.26]{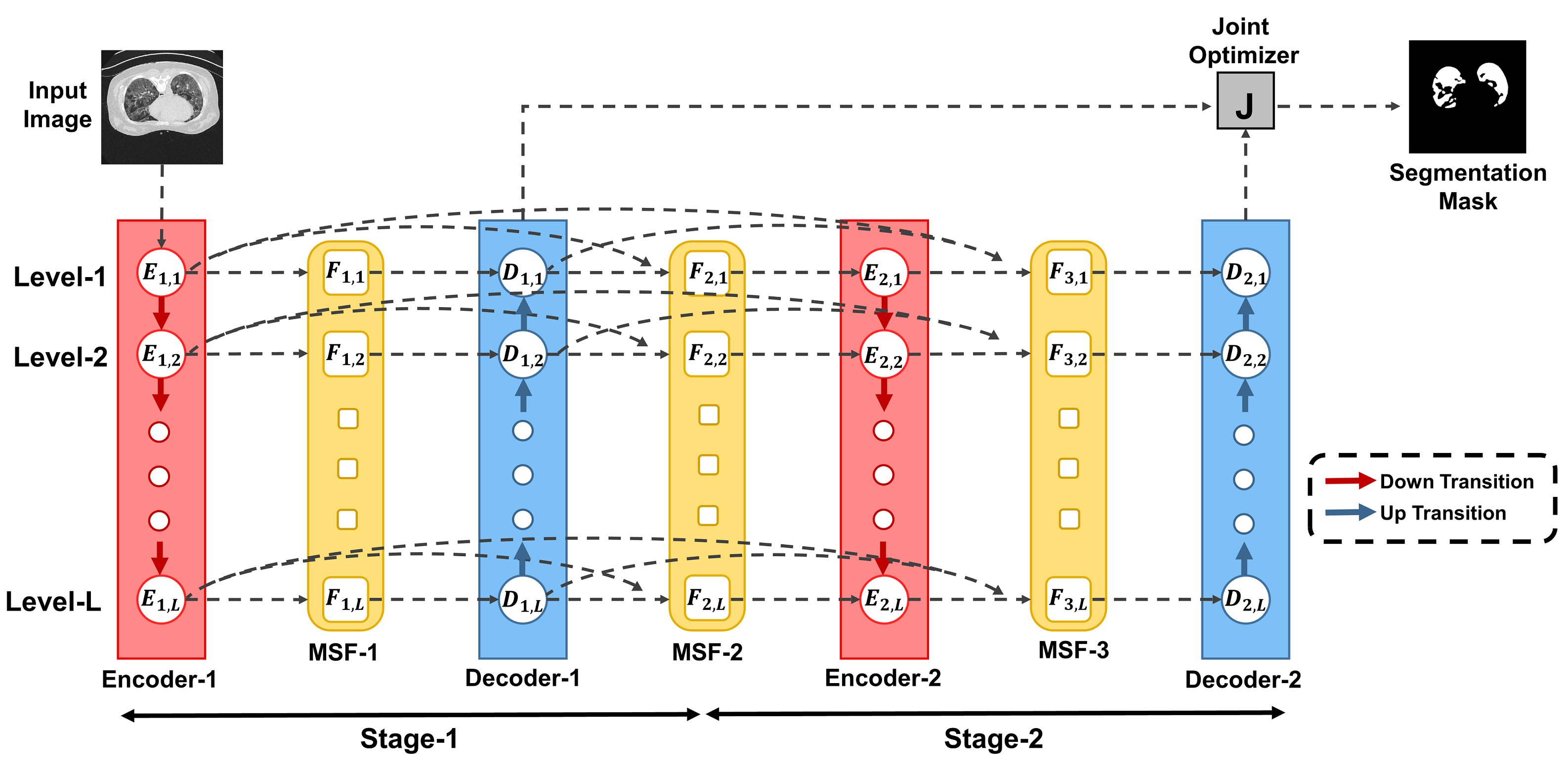}
    \caption{\textbf{Schematic representation of the two-stage implementation of the proposed CovSegNet architecture where two sequential encoder-decoder operational stages are employed with $L$ subsequent levels. Three multi-scale fusion (MSF) modules are integrated in between subsequent modules. Generated feature maps from two decoder modules are optimized using the joint optimizer. These encoder, decoder and MSF modules are composed of several operational unit cells.}}
    \label{f2}
\end{figure*}

\section{Methodology}

The proposed scheme splits the segmentation of CT volumes into two subsequent phases to reduce the computational complexity of 3D convolution as well as to take the advantages of multi-scale 2D convolutions (Fig.~\ref{f1}). In the first phase of the training, 2D slice-based optimization process is carried out where a 2D variant of the proposed CovSegNet architecture (i.e. CovSegNet2D) is employed to extract the segmentation mask of the infected lesions in CT slices. After optimization, a thresholding scheme is employed to convert the predicted probability mask into a binary mask. Hence, after completion of the phase-1 of training and optimization, this network is capable of extracting slice-based lesion mask efficiently and effectively. However, slice-based processing of input CT volumes will lead to loss of inter-slice contextual information resulting in sub-optimal performance.
To introduce further optimization and processing utilizing the inter-slice information, phase-2 of the training stage is incorporated. Several 2D-slices are extracted from input CT volumes and the pre-trained CovSegNet2D is utilized to extract the probability masks of the lung lesions. As CovSegNet2D is heavily optimized in phase-1 for 2D-slice based segmentation, it will provide the effective probability mask of the region-of-interest (ROI) in the CT slices. These masks are used for enhancing the ROIs of the CT slices while suppressing the redundant parts, and these are aggregated later to generate the ROI enhanced CT volume where most of the redundant parts are removed. Afterwards, the 3D variant of the proposed CovSegNet (i.e. CovSegNet3D) is employed into operation for further processing of the ROI enhanced CT volume considering both intra-slice and inter-slice contextual features. 
At the phase-2 of training, this CovSegNet3D is trained and optimized for generating the 3D-volumetric probability mask to introduce the inter-slice processing for improving performance while the pre-trained CovSegNet2D obtained from phase-1 is supposed to be fine-tuned for generating ROI-enhanced slices. Both these networks pass through a joint-optimization process for achieving optimum performance.
Moreover, a deeper variant of CovSegNet2D is used for exploiting the advantages of less expensive 2D operations while a shallower variant of CovSegNet3D is used to reduce the computational burden of 3D processing. As considerably precise performance can be achieved from the slice-based operations utilizing the CovSegNet2D only, it minimizes the need for deeper 3D-operations in phase-2 of training. Hence, the proposed hybrid networking scheme is capable of exploiting the advantages of both the efficient, lighter 2D convolutions along with the 3D contextual information that provides optimal performance.

%the main purpose of the CovSegNet3D is to provide some improved optimization from the enhanced ROI masks

%In this scheme, the main purpose of the CovSegNet3D is to provide some improved optimization from the enhanced ROI masks. Hence, a lightweight 3D-variants of CovSegNet is capable of achieving optimum performance as it is operated on the closely optimal output generated from 2D-counterparts. 

\subsection{Problem Formulation}
Let consider the set of CT volumes as $\mathbf{X}$, and their corresponding ground truths as $\mathbf{Y}$, such that $X_i \in \R^{h\times w \times s \times c}$, $Y_i \in \R^{h\times w \times s \times c}$, and $i=\{1,2,3,\dots, N\}$, where $(h,w,s,c)$ denote height, width, number of slices, and channels per slice, respectively, of a particular CT volume from total $N$ number of CT volumes. Moreover, let consider $\mathbf{x_{i,j}} \in R^{h\times w \times c}$ as the $i_{th}$ slice from total $S$ slices of $j_{th}$ CT volume and $\mathbf{y_{i,j}}\in R^{h\times w\times c}$ as its corresponding mask, such that $i=\{1,2,\dots, S\}$, and $j=\{1,2,\dots, N\}$. In the first phase of training, the objective function for slice-based optimization of CovSegNet2D is

\begin{equation}
    \centering
    \textbf{Phase 1: }\argmin_{\theta} \mathscr{L}_{2D}(\theta,\ \mathbf{y^{p},\ y})
\end{equation}
where, $\theta$ denotes the network parameter of CovSegNet2D, $\mathbf{x, y^p, y}$ denote the input 2D-slice, predicted probability mask, and corresponding ground truth mask.

% \usepackage{multirow}
\begin{table*}[t]
\centering
\caption{Architectural and Operational Details of the Encoder, Decoder, and Multi-Scale Fusion Modules of the Proposed CovSegNet2D for Optimum Performance in Independent Single-Network Implementation}
\label{t1}
\scalebox{0.8}{
\begin{tabular}{|c|c|c|c|c|c|c|c|c|}
\hline
\multicolumn{3}{|c|}{\textbf{Encoder Module}}                                                                                              & \multicolumn{3}{c|}{\textbf{Decoder Module}}                                                                                                 & \multicolumn{3}{c|}{\textbf{Multi-Scale Fusion Module}}                                                                                                                              \\ \hline
\textbf{Unit}         & \textbf{Ingredients}                                                                 & \textbf{Output}             & \textbf{Unit}         & \textbf{Ingredients}                                                                   & \textbf{Output}             & \textbf{Unit}          & \textbf{Ingredients}                                                                                                          & \textbf{Output}             \\ \hline
E-1                   & (Conv 1$\times$1, Conv 3$\times$3) $\times$  4                                                            & 512$\times$512$\times$16                  & D-5                   & (Conv 1$\times$1, Conv 3$\times$3) $\times$  4                                                              & 32$\times$32$\times$256                   & \multirow{3}{*}{MSF-1} & \multirow{3}{*}{\begin{tabular}[c]{@{}c@{}}Upsample(2$\times$2,4$\times$4,8$\times$8,16$\times$16)\\ Maxpool(2$\times$2,4$\times$4)\\ Conv 1$\times$1, Conv 3$\times$3\end{tabular}}  & \multirow{3}{*}{512$\times$512$\times$16} \\ \cline{1-6}
\multirow{2}{*}{DT-1} & \multirow{2}{*}{Conv 2$\times$2, Stride 2}                                                  & \multirow{2}{*}{256$\times$256$\times$32} & \multirow{2}{*}{UT-4} & \multirow{2}{*}{Deconv 2$\times$2, Stride 2}                                                  & \multirow{2}{*}{64$\times$64$\times$128}  &                        &                                                                                                                               &                             \\
                      &                                                                                      &                             &                       &                                                                                        &                             &                        &                                                                                                                               &                             \\ \hline
E-2                   & (Conv 1$\times$1, Conv 3$\times$3) $\times$  4                                                            & 256$\times$256$\times$32                  & D-4                   & (Conv 1$\times$1, Conv 3$\times$3) $\times$  4                                                              & 64$\times$64$\times$128                   & \multirow{2}{*}{MSF-2} & \multirow{2}{*}{\begin{tabular}[c]{@{}c@{}}Maxpool (2$\times$2,4$\times$4)\\ Upsample(2$\times$2,4$\times$4,8$\times$8)\\ Conv 1$\times$1, Conv 3$\times$3\end{tabular}}       & \multirow{2}{*}{256$\times$256$\times$32} \\ \cline{1-6}
DT-2                  & \begin{tabular}[c]{@{}c@{}}Maxpool 2$\times$2\\ Conv 2$\times$2, Stride 2\end{tabular}             & 128$\times$128$\times$64                  & UT-3                  & \begin{tabular}[c]{@{}c@{}}Upsample 2$\times$2\\ Deconv 2$\times$2, Stride 2\end{tabular}            & 128$\times$128$\times$64                  &                        &                                                                                                                               &                             \\ \hline
E-3                   & (Conv 1$\times$1, Conv 3$\times$3) $\times$  4                                                            & 128$\times$128$\times$64                  & D-3                   & (Conv 1$\times$1, Conv 3$\times$3) $\times$  4                                                              & 128$\times$128$\times$64                  & \multirow{2}{*}{MSF-3} & \multirow{2}{*}{\begin{tabular}[c]{@{}c@{}}Maxpool (2$\times$2,4$\times$4)\\ Upsample(2$\times$2,4$\times$)\\ Conv 1$\times$1, Conv 3$\times$3\end{tabular}}           & \multirow{2}{*}{128$\times$128$\times$64} \\ \cline{1-6}
DT-3                  & \begin{tabular}[c]{@{}c@{}}Maxpool (2$\times$2, 4$\times$4)\\ Conv 2$\times$2, Stride 2\end{tabular}      & 64$\times$64$\times$128                   & UT-2                  & \begin{tabular}[c]{@{}c@{}}Upsample (2$\times$2, 4$\times$4)\\ Deconv 2$\times$2, Stride 2\end{tabular}     & 256$\times$256$\times$32                  &                        &                                                                                                                               &                             \\ \hline
E-4                   & (Conv 1$\times$1, Conv 3$\times$3) $\times$  4                                                            & 64$\times$64$\times$128                   & D-2                   & (Conv 1$\times$1, Conv 3$\times$3) $\times$  4                                                              & 256$\times$256$\times$32                  & \multirow{2}{*}{MSF-4} & \multirow{2}{*}{\begin{tabular}[c]{@{}c@{}}Maxpool (2$\times$2,4$\times$4,8$\times$8)\\ Upsample(2$\times$2,4$\times$4)\\ Conv 1$\times$1, Conv 3$\times$3\end{tabular}}       & \multirow{2}{*}{64$\times$64$\times$128}  \\ \cline{1-6}
DT-4                  & \begin{tabular}[c]{@{}c@{}}Maxpool (2$\times$2, 4$\times$4, 8$\times$8)\\ Conv 2$\times$2, Stride 2\end{tabular} & 32$\times$32$\times$256                   & UT-1                  & \begin{tabular}[c]{@{}c@{}}Maxpool (2$\times$2, 4$\times$4, 8$\times$8)\\ Deconv 2$\times$2, Stride 2\end{tabular} & 512$\times$512$\times$16                  &                        &                                                                                                                               &                             \\ \hline
\multirow{3}{*}{E-5}  & \multirow{3}{*}{(Conv 1$\times$1, Conv 3$\times$3) $\times$  4}                                           & \multirow{3}{*}{32$\times$32$\times$256}  & \multirow{3}{*}{D-1}  & \multirow{3}{*}{(Conv 1$\times$1, Conv 3$\times$3) $\times$  4}                                             & \multirow{3}{*}{512$\times$512$\times$16} & \multirow{3}{*}{MSF-5} & \multirow{3}{*}{\begin{tabular}[c]{@{}c@{}}Maxpool(2$\times$2,4$\times$4,8$\times$8,16$\times$16)\\ Upsample(2$\times$2, 4$\times$4)\\ Conv 1$\times$1, Conv 3$\times$3\end{tabular}} & \multirow{3}{*}{32$\times$32$\times$256}  \\
                      &                                                                                      &                             &                       &                                                                                        &                             &                        &                                                                                                                               &                             \\
                      &                                                                                      &                             &                       &                                                                                        &                             &                        &                                                                                                                               &                             \\ \hline
\end{tabular}
}
\end{table*}

% \usepackage{multirow}
\begin{table*}[t]
\centering
\caption{Architectural and Operational Details of the Encoder, Decoder, and Multi-Scale Fusion Modules of the Proposed CovSegNet3D for Optimum Performance in Independent Single-Network Implementation}
\label{t2}
\scalebox{0.7}{
\begin{tabular}{|c|c|c|c|c|c|c|c|c|}
\hline
\multicolumn{3}{|c|}{\textbf{Encoder Module}}                                                                  & \multicolumn{3}{c|}{\textbf{Decoder Module}}                                                                                     & \multicolumn{3}{c|}{\textbf{Multi-Scale Fusion Module}}                                                                                                                                         \\ \hline
\textbf{Unit} & \textbf{Ingredients}                                                         & \textbf{Output} & \textbf{Unit} & \textbf{Ingredients}                                                                           & \textbf{Output} & \textbf{Unit}          & \textbf{Ingredients}                                                                                                                  & \textbf{Output}                \\ \hline
E-1           & (Conv 1$\times$1$\times$1, Conv 3$\times$3$\times$3) $\times$  2                                                & 512$\times$512$\times$32$\times$16   & D-4           & (Conv 1$\times$1$\times$1, Conv 3$\times$3$\times$3) $\times$ 2                                                                   & 64$\times$64$\times$4$\times$128     & \multirow{2}{*}{MSF-1} & \multirow{2}{*}{\begin{tabular}[c]{@{}c@{}}Maxpool(2$\times$2$\times$2,4$\times$4$\times$4)\\ Upsample(2$\times$2$\times$2, 4$\times$4$\times$4,8x8x8)\\ Conv 1$\times$1$\times$1, Conv 3$\times$3$\times$3\end{tabular}} & \multirow{2}{*}{512$\times$512$\times$32$\times$16} \\ \cline{1-6}
DT-1          & Conv 2$\times$2$\times$2, Stride 2                                                         & 256$\times$256$\times$16$\times$32   & UT-3          & \begin{tabular}[c]{@{}c@{}}Upsample (2$\times$2$\times$2, 4$\times$4$\times$4)\\ Deconv 2$\times$2$\times$2, Stride 2\end{tabular}       & 128$\times$128$\times$8$\times$64    &                        &                                                                                                                                       &                                \\ \hline
E-2           & (Conv 1$\times$1$\times$1, Conv 3$\times$3$\times$3) $\times$  2                                                & 256$\times$256$\times$16$\times$32   & D-3           & (Conv 1$\times$1$\times$1, Conv 3$\times$3$\times$3) $\times$  2                                                                  & 128$\times$128$\times$8$\times$64    & \multirow{2}{*}{MSF-2} & \multirow{2}{*}{\begin{tabular}[c]{@{}c@{}}Maxpool (2$\times$2$\times$2, 4$\times$4$\times$4)\\ Upsample(2$\times$2$\times$2,4$\times$4$\times$4)\\ Conv 1$\times$1$\times$1, Conv 3$\times$3$\times$3\end{tabular}}      & \multirow{2}{*}{256$\times$256$\times$16$\times$32} \\ \cline{1-6}
DT-2          & \begin{tabular}[c]{@{}c@{}}Maxpool 2$\times$2$\times$2\\ Conv 2$\times$2$\times$2, Stride 2\end{tabular} & 128$\times$128$\times$8$\times$64    & UT-2          & \begin{tabular}[c]{@{}c@{}}Maxpool (2$\times$2$\times$2, 4$\times$4$\times$4, 8$\times$8$\times$8)\\ Deconv 2$\times$2$\times$2, Stride 2\end{tabular} & 256$\times$256$\times$16$\times$32   &                        &                                                                                                                                       &                                \\ \hline
E-3           & (Conv 1$\times$1$\times$1, Conv 3$\times$3$\times$3) $\times$  2                                                & 128$\times$128$\times$8$\times$64    & D-2           & (Conv 1$\times$1$\times$1, Conv 3$\times$3$\times$3) $\times$  2                                                                  & 256$\times$256$\times$16$\times$32   & \multirow{2}{*}{MSF-3} & \multirow{2}{*}{\begin{tabular}[c]{@{}c@{}}Maxpool (2$\times$2$\times$2, 4$\times$4$\times$4)\\ Upsample(2$\times$2$\times$2,4$\times$4$\times$4)\\ Conv 1$\times$1$\times$1, Conv 3$\times$3$\times$3\end{tabular}}      & \multirow{2}{*}{128$\times$128$\times$8$\times$64}  \\ \cline{1-6}
DT-3          & \begin{tabular}[c]{@{}c@{}}Maxpool 2$\times$2$\times$2\\ Conv 2$\times$2$\times$2, Stride 2\end{tabular} & 64$\times$64$\times$4$\times$128     & UT-1          & \begin{tabular}[c]{@{}c@{}}Maxpool (2$\times$2$\times$2, 4$\times$4$\times$4, 8$\times$8$\times$8)\\ Deconv 2$\times$2$\times$2, Stride 2\end{tabular} & 512$\times$512$\times$32$\times$16   &                        &                                                                                                                                       &                                \\ \hline
E-4           & (Conv 1$\times$1$\times$1, Conv 3$\times$3$\times$3) $\times$  2                                                & 64$\times$64$\times$4$\times$128     & D-1           & (Conv 1$\times$1$\times$1, Conv 3$\times$3$\times$3) $\times$  2                                                                  & 512$\times$512$\times$32$\times$16   & MSF-4                  & \begin{tabular}[c]{@{}c@{}}Maxpool (2$\times$2$\times$2,4$\times$4$\times$4)\\ Upsample(2$\times$2$\times$2,4$\times$4$\times$4, 8$\times$8$\times$8)\\ Conv 1$\times$1$\times$1, Conv 3$\times$3$\times$3\end{tabular}                 & 64$\times$64x4$\times$128                    \\ \hline
\end{tabular}
}
\end{table*}

In the phase-2 of training, the pre-trained CovSegNet2D network obtained from phase-1 is employed to generate ROI enhanced CT volume $\mathbf{X'}$, and thus
\begin{align}
    \centering
    &\mathbf{x'} = \mathbf{x} \odot \mathbf{y^{p}}\ ;\ \  \forall\ \mathbf{x'\in X',\ x\in X, \  y^p \in Y^p}
\end{align}
where $\odot$ denotes element-wise multiplication and $\mathbf{x}$ denotes 2D-CT slice, $\mathbf{x'}$ denotes ROI-enhanced CT-slice, and $\mathbf{y^p}$ denotes the predicted probability mask.
 
 Afterwards, optimization of the CovSegNet3D is carried out utlizing ROI-enhanced CT-volume, while CovSegNet2D is fine-tuned to generate more accurate probability masks from 2D-slices, and the joint optimization objective function $\mathcal{F}$ can be formulated as
 %using ROI-enhanced 3D CT volumes are carried out for CovSegNet3D that can be represented as
\begin{equation}
    \centering
    \textbf{Phase 2: }\argmin_{\Theta_1, \Theta_2} \mathcal{F}\{\mathscr{L}_{2D}(\Theta_1, \mathbf{y^p, y}),\mathscr{L}_{3D}(\Theta_2, \mathbf{Y^{p}, Y})\}
 \end{equation}
where $\Theta_1$ denotes the network parameters of CovSegNet2D, $\Theta_2$ denotes the network parameters of CovSegNet3D, $\mathbf{X', Y^p, Y}$ denote the ROI enhanced CT volume, predicted 3D mask, and corresponding 3D ground truth.

\subsection{Proposed CovSegNet architecture}
The proposed CovSegNet architecture is a generic representation of a network with a wide range of flexibility for increasing its applicability in different challenging conditions. This architecture can be designed for efficient operations in both 2D and 3D domains. Moreover, it can be made deeper/lighter according to the requirement of the applications. 

In CovSegNet architecture, multiple stages of sequential encoding and decoding operations are carried out along with a fusion scheme of multi-scale features in between subsequent encoder/decoder module. Each stage of the network consists of an encoder module and a corresponding decoder module. Hence, the network, $\mathcal{N}$, can be represented as
\begin{equation}
    \mathcal{N} = \mathbf{D_m(E_m\dots (D_1(E_1(\theta_{E_1}),\theta_{D_1}),\dots, \theta_{E_m}),\theta_{D_m})}
\end{equation}
where $\mathbf{E_i, D_i}$ represents the encoder and decoder modules, respectively, of $i_{th}$ stage from total $m$ stages, and $\theta_{E_i}, \theta_{D_i}$ represents their respective parameters.
Two-stage implementation of this architecture is schematically presented in Fig.~\ref{f2}.

This network can be extended from level-1 to level-L to produce a deeper variant. 
The encoder/decoder module constitutes of several unit cells operating at each level of the network. To generate a deeper network, additional unit cells are integrated in each of the encoder/decoder module to increase number of levels. Here, $E_{i,j}$, $D_{i,j}$ represent the $i_{th}$ unit cell of $j_{th}$ stage of encoder and decoder, respectively, where $i= \{1,2,\dots,L\}$, and $j=\{1,2, \dots, m\}$.
Hence, $L$ number of different scales of representative feature maps are obtained from each encoder/decoder module. Moreover, scale transition of feature maps is carried out in between succeeding encoder/decoder unit cells, and effective transformation on each scale of feature maps are integrated utilizing the generalized unit cell structure in encoder/decoder module.

In between successive encoder/decoder modules, a multi-scale fusion (MSF) module is introduced to reduce the semantic gap with preceding stages as well as to improve the gradient propagation through parallel linkage of multi-scale features. Similar to encoder/decoder module, each MSF module consists of several operational unit cells operating at different levels. Let consider, $\mathbf{F_i}$ represents the $i_{th}$ MSF module, $F_{i,j}$ represents the $i_{th}$ unit cell of $j_{th}$ MSF module, such that $i=\{1,2,\dots, L\}$, $j=\{1,2,\dots,2m-1\}$, and $F_{i,j}\in \mathbf{F_i}$.  

Each MSF module takes all scales of feature representations as input from all preceding encoder/decoder stages, and generates $L$ number of different feature maps for the following encoder/decoder stage through deep fusion of multi-scale features obtained from preceding stages. In each unit cell of MSF module, multi-scale feature aggregation and pyramid fusion scheme is employed, which can be represented as
\begin{align}
    &F_{i,j} = \mathscr{F}(\mathbf{E_1, E_2,\dots, E_{\ceil{\frac{j}{2}}}, D_{1}, D_2, \dots, D_{\floor{\frac{j}{2}}}}) \\
   &\nonumber \forall \ i=\{1, 2, \dots, L\},\ j=\{1, 2,\dots, 2m-1\}
\end{align}
where $\mathscr{F}(.)$ represents the functional operations in the MSF unit cell considering $L$ scale of representations from each of the preceding encoder/decoder module.

From final level of the sequential decoder modules, several decoded feature representations are obtained which are processed together in the joint optimizer unit ($\mathcal{J}$) to produce the final segmentation mask, and it can be given by,
\begin{equation}
    \mathcal{J} = \mathcal{F}(D_{1,1},D_{1,2}, \dots, D_{1,m})
\end{equation}
where $\mathcal{F}(.)$ represents the joint optimizer function.

All the basic building blocks of the CovSegNet architecture are generic and can be designed and optimized for both 2D and 3D operations. In the following discussions, different building blocks of the CovSegNet architecture are presented in detail. For the ease of discussion, mainly 2D operational blocks are focused. However, for 3D operations, all the convolutional kernels, pooling/upsampling windows are shifted in dimension for operating with 3D voxels instead of 2D pixels. Architectural details for the most optimized implementations of CovSegNet2D and CovSegNet3D are presented in Table~\ref{t1} and in Table~\ref{t2}, respectively.

\begin{figure}[!t]
    \centering
    \subfloat[\textbf{Encoder Module}]{\includegraphics[scale=0.12]{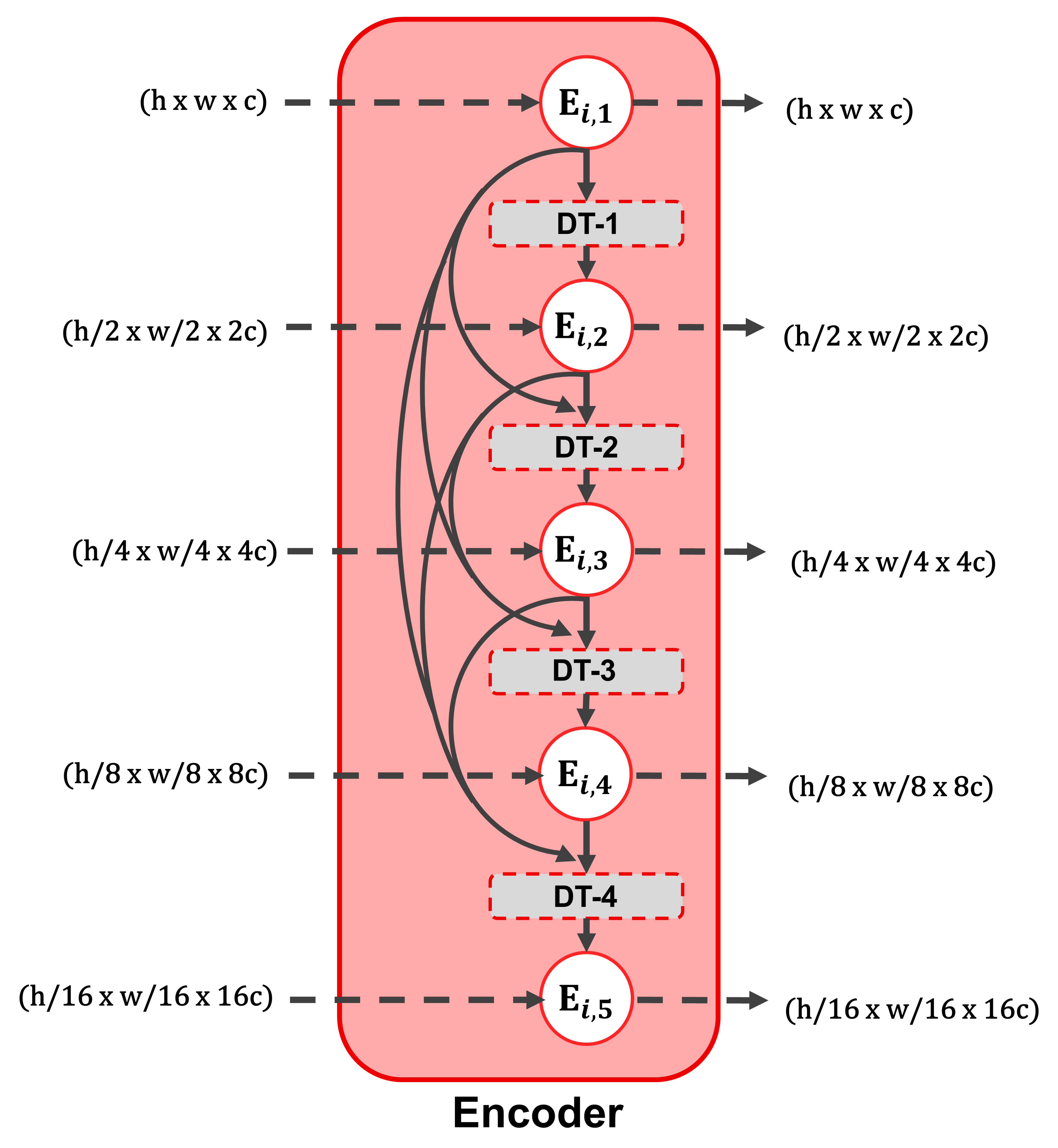}%  
    \label{f3a}}
    \hspace{2mm}
    \subfloat[\textbf{Decoder Module}]{\includegraphics[scale=0.12]{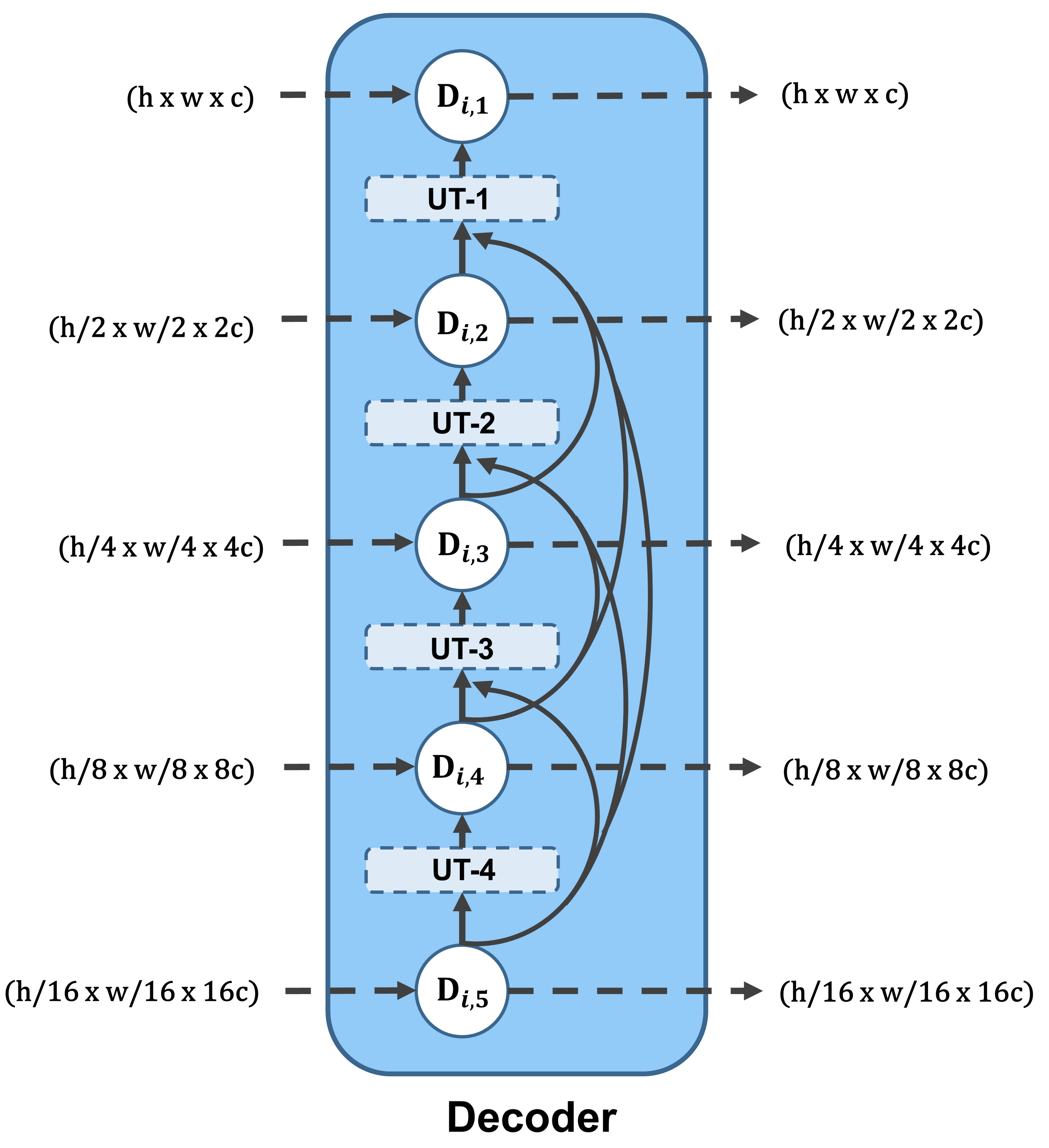}%
    \label{f3b}}
    \caption{\textbf{Schematic representations of the proposed encoder and decoder modules in five-level implementation having five unit blocks along with associated Down Transition (DT)/ Up Transition (UT) units in between subsequent unit blocks. Here, $(h, w, c)$  is used to denote the height, width, and channel of the feature maps at different phase.}} 
    \label{f3}
\end{figure}

\begin{figure}[!t]
    \centering
    \subfloat[\textbf{Encoder Unit Cell}]{\includegraphics[scale=.15]{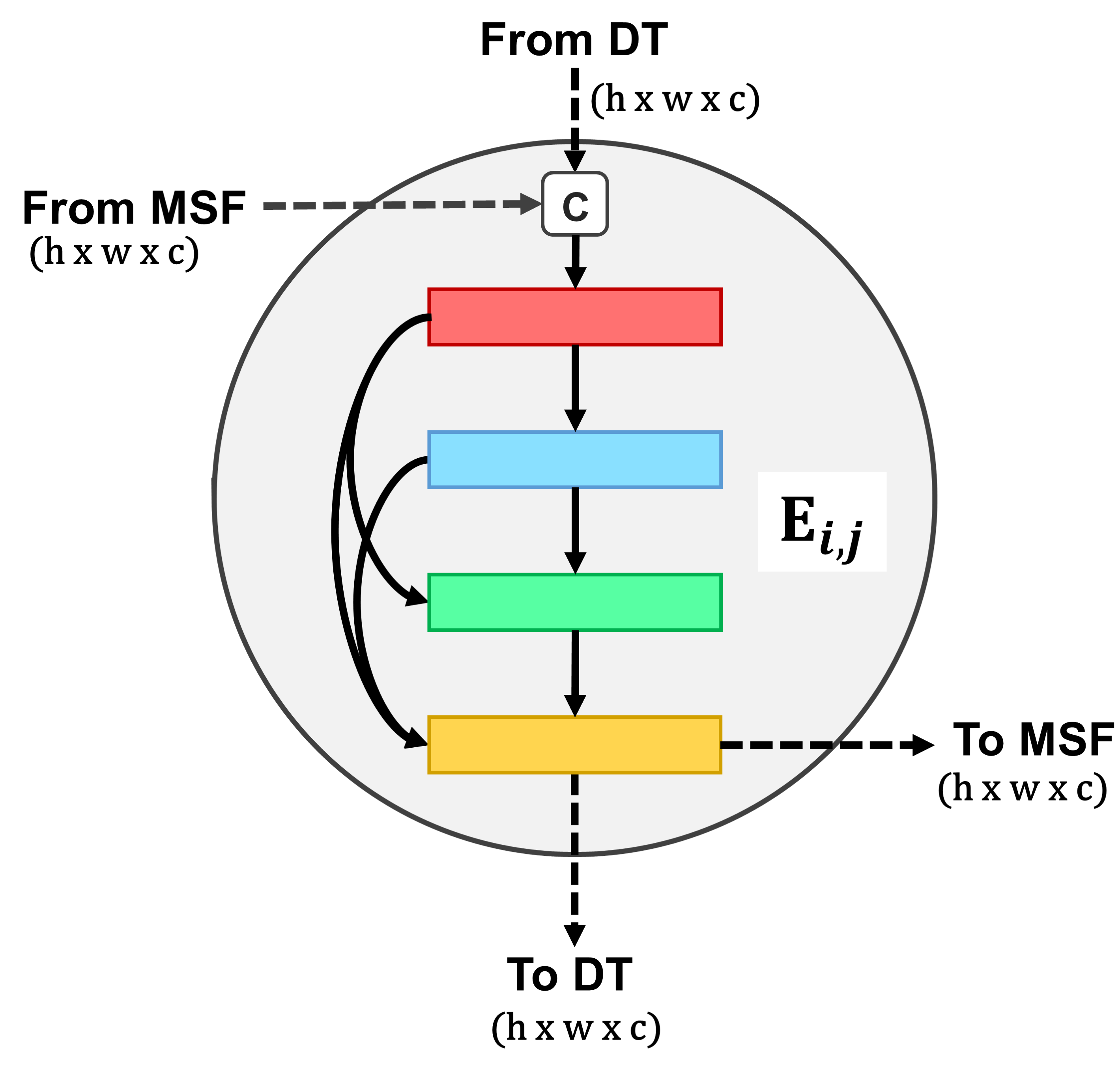}%  
    \label{f4a}}
    \hspace{1cm}
    \subfloat[\textbf{Decoder Unit Cell}]{\includegraphics[scale=.15]{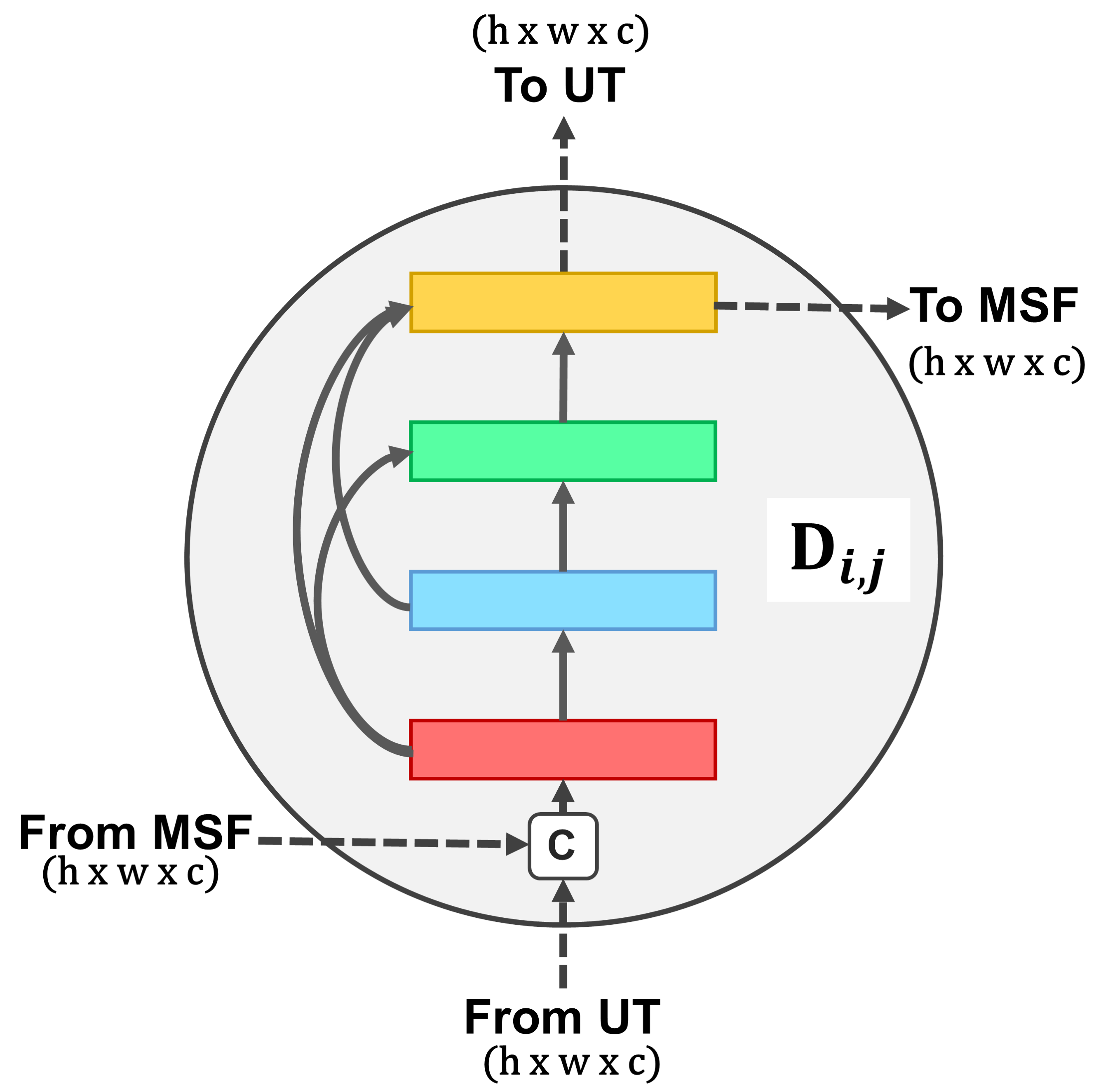}%
    \label{f4b}}
    \caption{\textbf{Structure of the Encoder/Decoder Unit cells. Four densely interconnected convolutional stages are employed in each unit. Here, `c' denotes the channelwise concatenation of feature maps extracted from transition unit and multi-scale fusion (MSF) unit. $E_{i,j}/D_{i,j}$ denote the unit blocks of $i_{th}$ level in $j_{th}$ module.}} 
    \label{f4}
\end{figure}

\begin{figure}[!t]
    \centering
    \subfloat[\textbf{Down Transition Unit (DT-3)}]{\includegraphics[scale=.26]{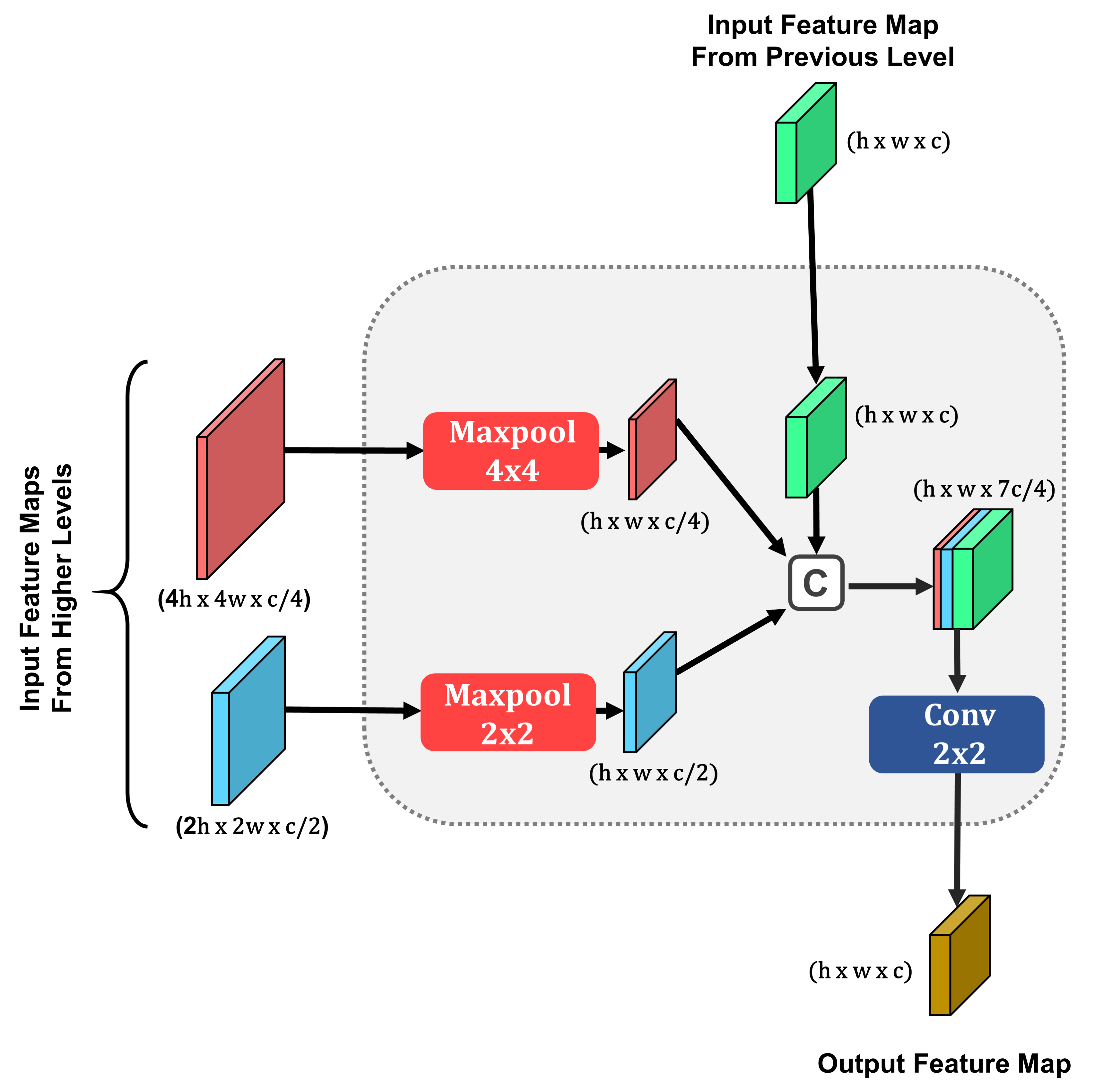}%  
    \label{f5a}}
%    \hspace{1.3cm}
    \vfill
    \subfloat[\textbf{Up Transition Unit (UT-2)}]{\includegraphics[scale=.26]{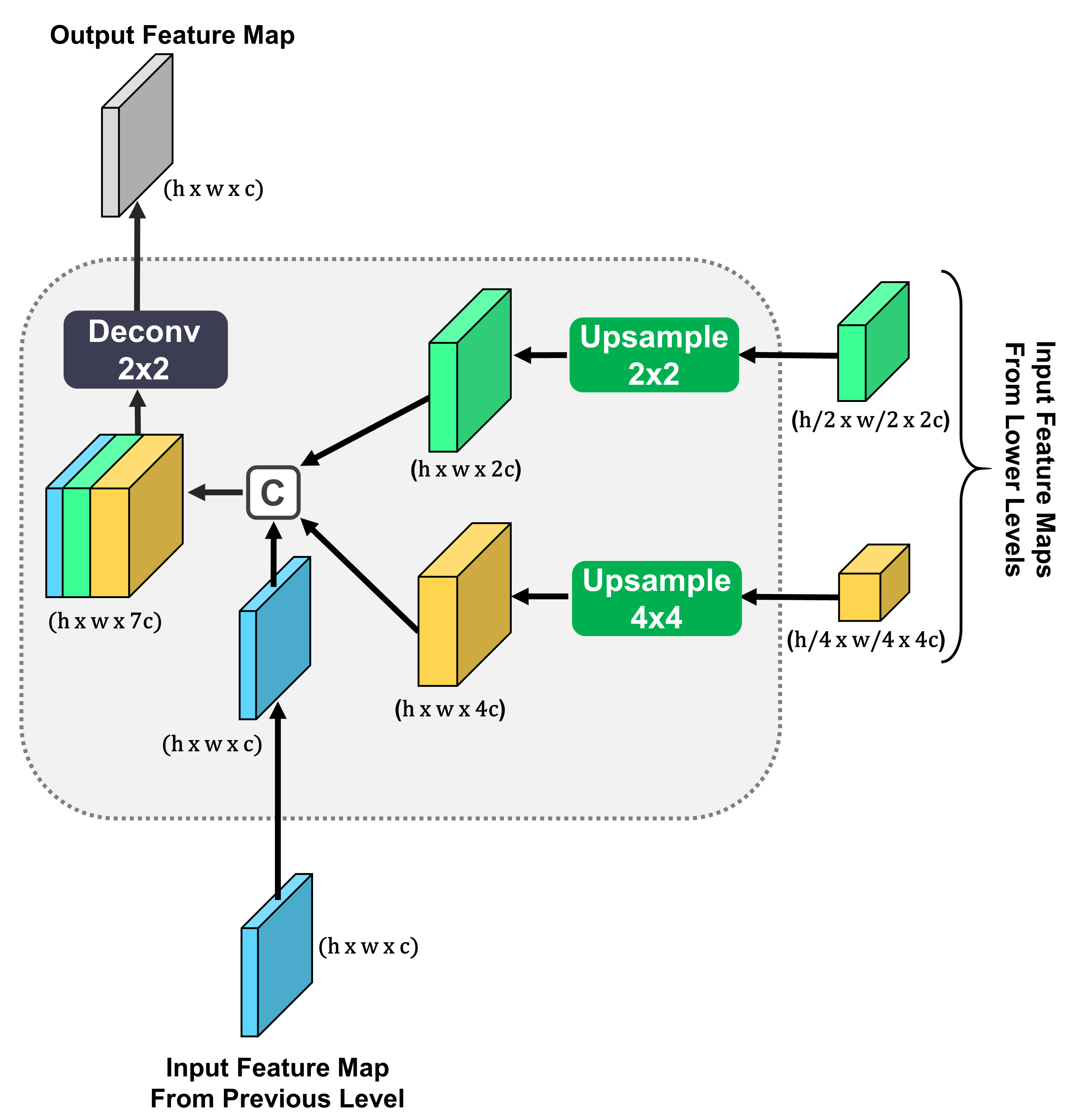}%
    \label{f5b}}
    \caption{\textbf{Schematic representations of the down transition unit (operating between level-3 and 4) and the up transition unit (operating between level-($L-2$) and ($L-3$)). All the feature maps generated from preceding unit blocks are made uniform and integrated in the transition process.}} 
    \label{f5}
\end{figure}

\subsection{Proposed Encoder/Decoder Structure}
The encoder and decoder modules are structurally similar that are successively used in the sequential stages of CovSegNet. Encoder/decoder modules are schematically presented in Fig.~\ref{f3}. These encoder/decoder modules are composed of several operational unit cells with transitional dense interconnections. The operations of encoder/decoder modules can be divided into two categories: unit cell operations and transitional operation.

\subsubsection{Encoder/Decoder Unit Cell operation}
In Fig.~\ref{f4}, the unit cell structure of the encoder/decoder module is presented. In each unit cell, two input feature map is entered, one from the transitional unit and the other from the preceding MSF unit while the output feature map is passed through following transitional and multi-scale fusion operations. Moreover, each unit cell consists of four densely interconnected convolutional layers, where each convolutional layer provides two sequential convolutional filtering with $(1\times 1)$ and $(3\times 3)$ kernels. Such dense interconnection between convolutional operations has been proven to be effective in numerous applications. No dimensional scaling has been carried out in each of this unit cell as it is employed for introducing adequate transformation in the feature space to encode/decode effective representation. Hence, this unit cell operations can be functionally represented as, $E,D: \R^{h\times w \times c} \rightarrow \R^{h\times w \times c}$, where $(h, w, c)$ represents the height, width and channel of the feature map.

\begin{figure}[t]
    \centering
    \includegraphics[scale=0.12]{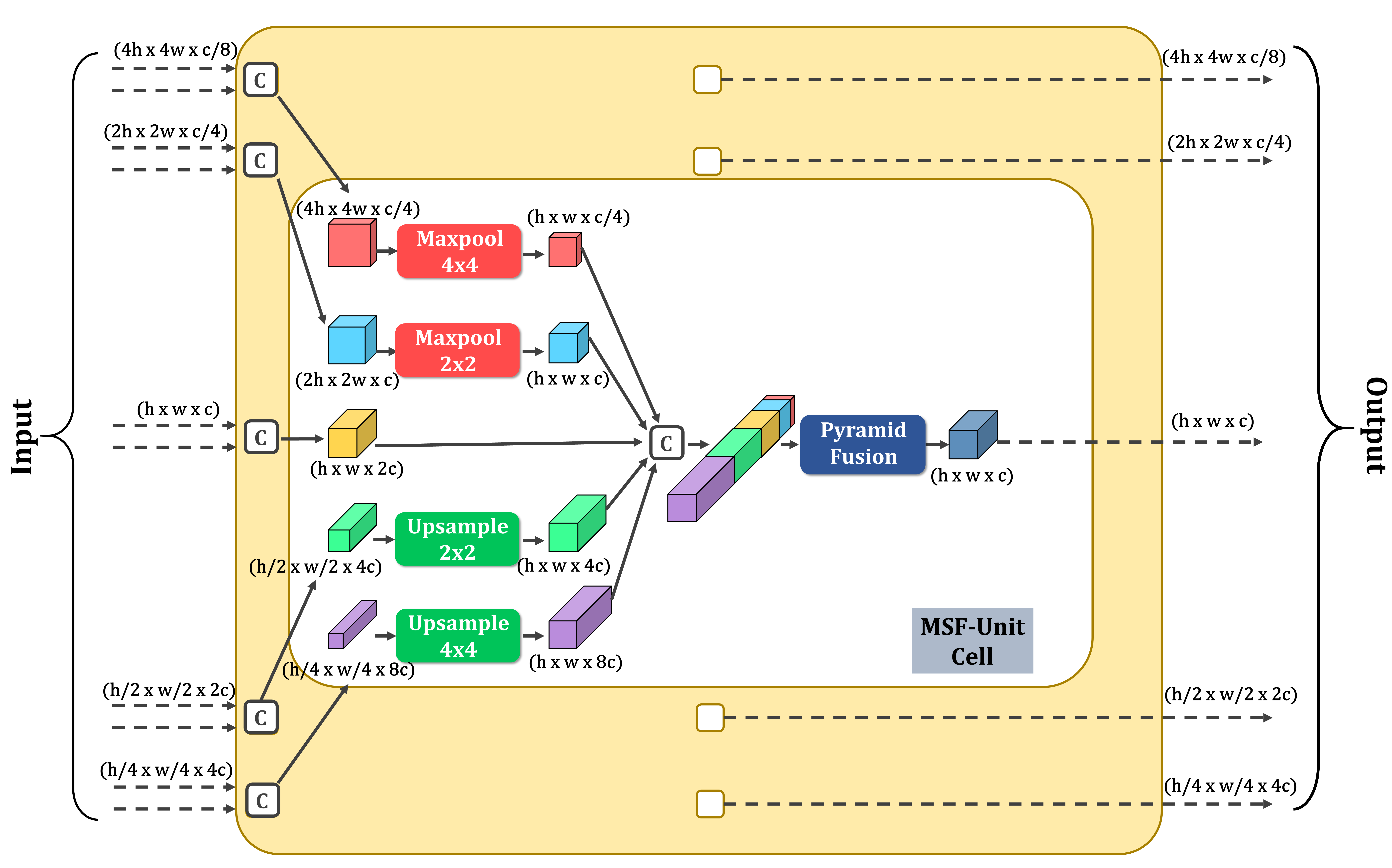}
    \caption{\textbf{Schematic representation of the proposed Multi-Scale Fusion module. Detailed operations performed in an MSF unit cell are particularly focused, and similar operations are carried out in other unit cells of the MSF module.}}
    \label{f6}
\end{figure}

\begin{figure}[t]
    \centering
    \includegraphics[scale=0.18]{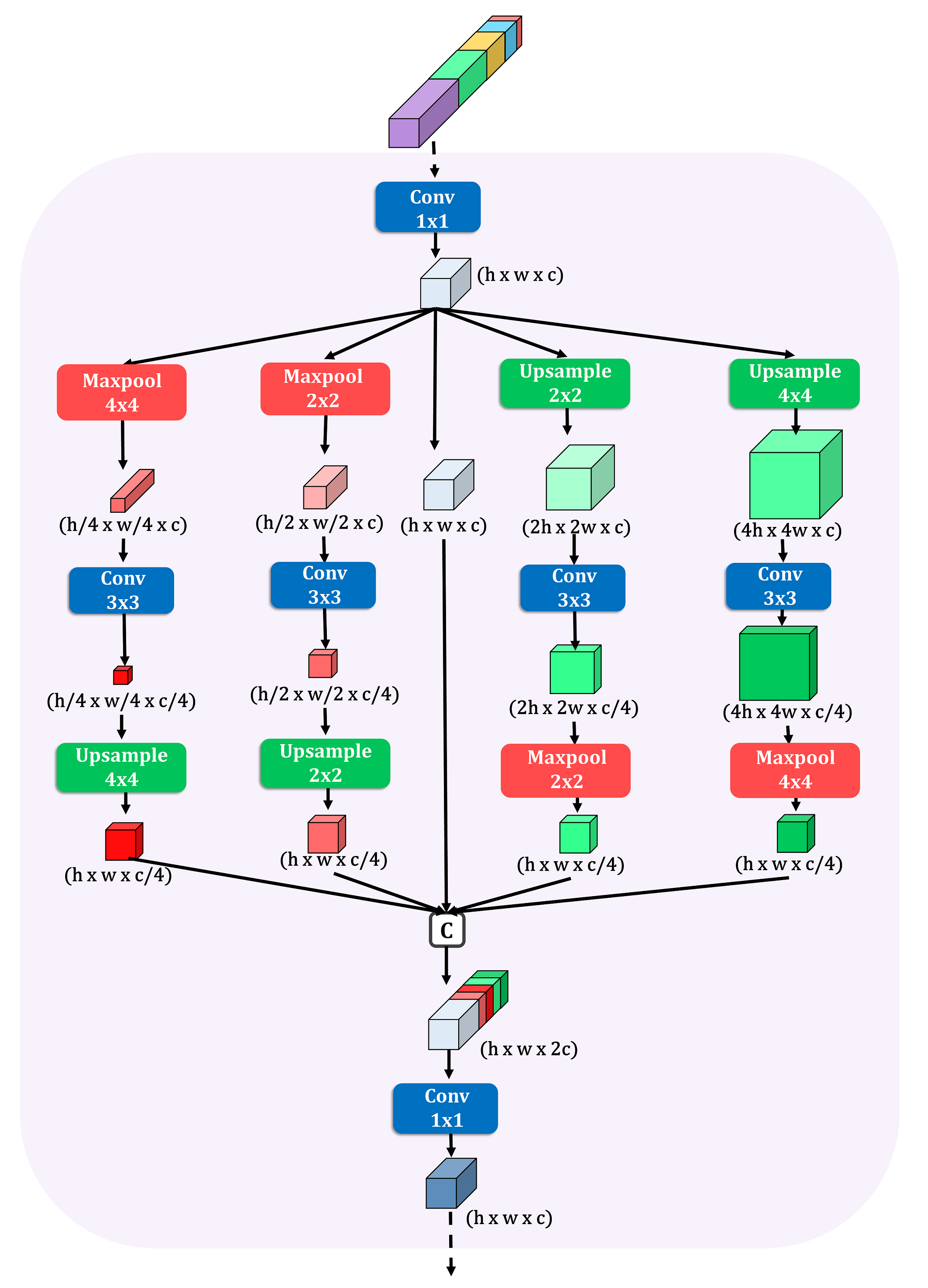}
    \caption{\textbf{Proposed pyramid fusion scheme utilizing diverse windows of frequent up-sampling and downsampling operations for fusing multi-scale features.}}
    \label{f7}
\end{figure}

\begin{figure}[t]
    \centering
    \includegraphics[scale=0.17]{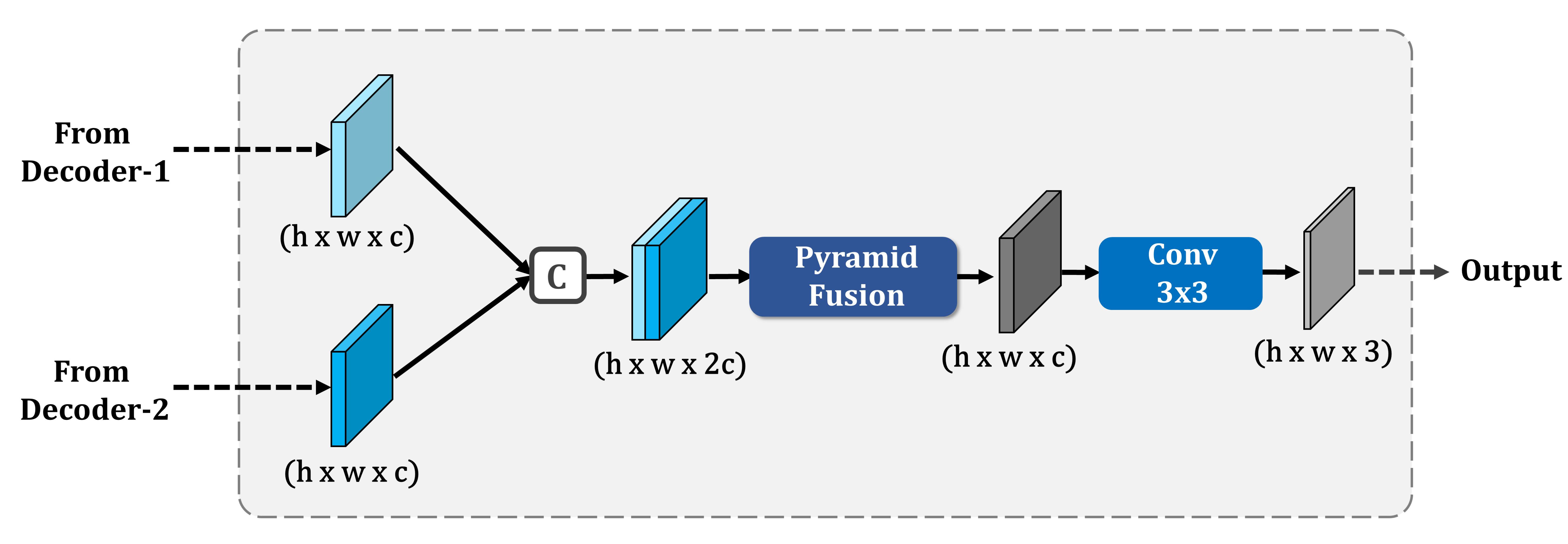}
    \caption{\textbf{Schematic of the joint optimizer module optimizing the decoded feature maps generated from two decoding stages.}}
    \label{f8}
\end{figure}

\subsubsection{Encoder Down-transitional Operation}
During down transitional operations between subsequent unit cells of the encoder module, the spatial dimension of the feature map is reduced for generalizing the feature map, whereas the channel depth is increased to incorporate more filtering operations in subsequent levels for generating more sparser features. It can be functionally presented as, $f: \R^{h\times w \times c} \rightarrow \R^{h/2\times w/2 \times 2c}$, where spatial resolution is downscaled by 2 and channel depth is increased by 2 from the input feature map obtained from the previous level. However, traditional downsampling operations using pooling/strided convolutions results in loss of contextual information. Moreover, it can be more prominent while incorporating a deep stack of unit cells in the encoder module. To mitigate the loss of contextual information in down transitional operation, a higher level of dense interconnection is proposed among multi-scale feature maps generated from different unit cells. In Fig.~\ref{f5a}, the structure of such a down transition unit is schematically presented. In each of such down transition unit, encoded feature representations generated from all higher levels of unit cells are considered for generating the down-scaled feature map. 
Hence, contextual information lost in each transitional operation can be recovered from very deep stack of unit cells as feature representations from all preceding cells are considered during transition. To converge multi-scale feature maps from preceding levels, firstly, pooling operations with different kernels are carried out to make their spatial dimension uniform and subsequently, channelwise feature aggregation is carried out. The aggregated feature map, $F_{agg, DT}$, generated at $i_{th}$ level can be represented as
\begin{equation}
    F_{agg, DT}^i = E^{i} \oplus P^{(2\times 2)}(E^{i-1}) \dots \oplus P^{(2^{i-1} \times 2^{i-1})} (E^{1})
\end{equation}
where $\oplus$ indicates the feature concatenation, $P^{(2\times2)}$ represents pooling operation with $(2\times 2)$ window, $E^i$ represents the output of $i_{th}$ unit cell of the encoder.

Finally, a convolutional operation with $(2\times 2)$ kernel is carried out with a stride of $(2\times 2)$ for generating the downscaled feature map by filtering the aggregated feature vector.

\begin{figure*}[t]
    \centering
    \includegraphics[scale=0.35]{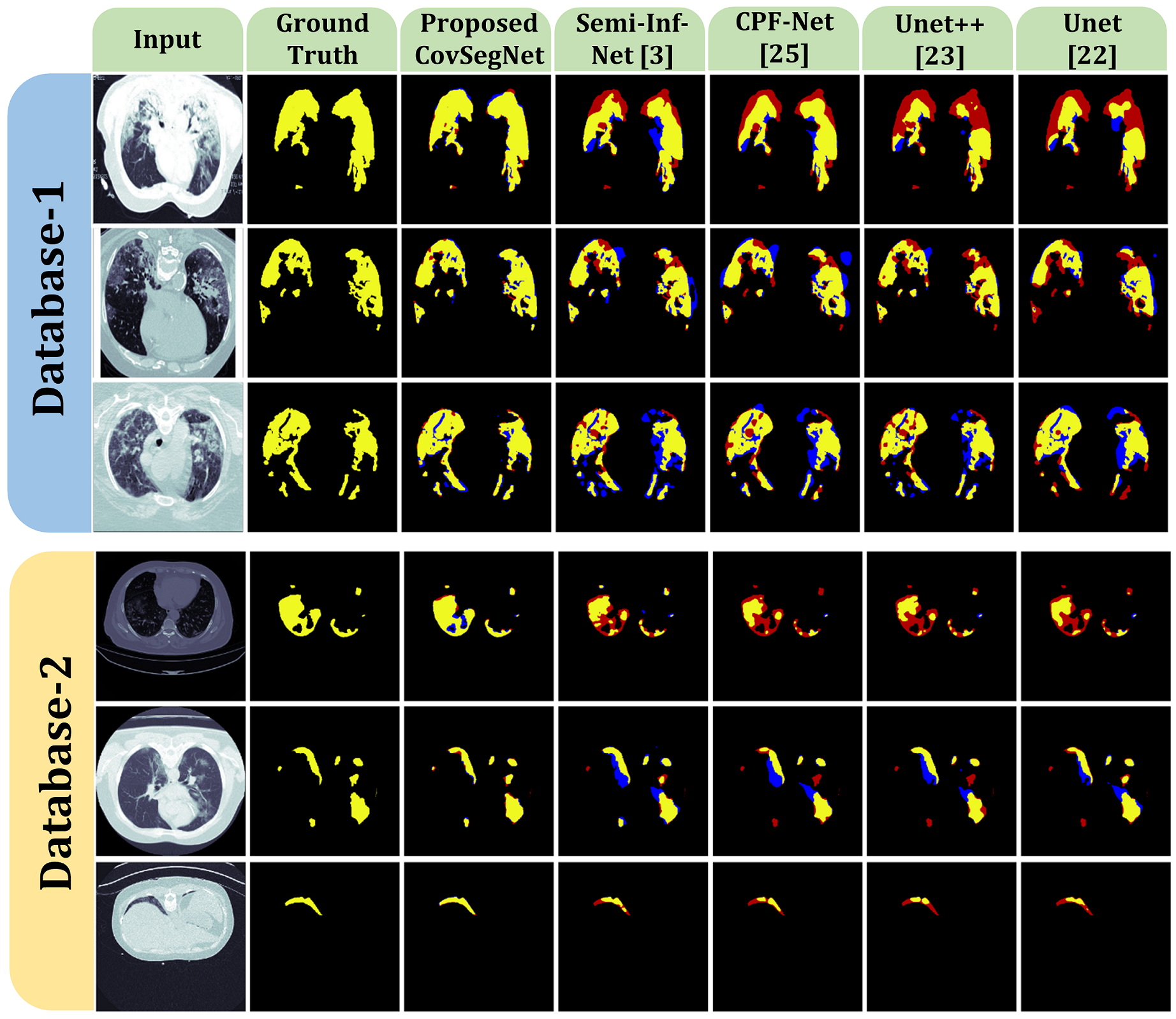}
    \caption{\textbf{Visual representations of the segmentation performances of different state-of-the-art networks on the CT images from Database-1 and Database-2. Here, `yellow' represents the true positive (TP) regions, `red' represents the false negative (FN) regions, and `blue' represents the false positive (FP) regions.}}
    \label{f9}
\end{figure*}

\subsubsection{Decoder Up-transitional Operation}
On the contrary, up transitional operations are carried out in between successive decoder unit cells to provide the dimensional shifting towards the reconstruction of the final segmentation mask. In each of such up-transition operations, spatial resolution is upscaled by 2 while channel depth is reduced by 2 to get closer to the final reconstruction mask and it can be represented as, $f':\R^{h\times w\times c} \rightarrow \R^{2h\times 2w \times c/2}$. Similar to the down-transitional operation in Encoder, all the preceding representations of multi-scale decoded feature maps generated from different unit cells are taken into consideration in the up-transition operation to gather more contextual information (Fig.~\ref{f5b}). Firstly, spatially uniform feature maps are created through bi-linear interpolation upsampling with different windows, and feature aggregation is carried out to generate aggregated feature vector, $F_{agg, UT}$, which is given by
\begin{equation}
    F_{agg, UT}^i = D^{i} \oplus U^{(2\times 2)}(D^{i+1}) \dots \oplus U^{(2^{i-1} \times 2^{i-1})} (D^{L})
\end{equation}
where $U^{(2\times2)}$ represents bilinear upsampling operation with $(2\times 2)$ window, $D^i$ represents the output of $i_{th}$ unit cell of the decoder.

Finally, the aggregated feature map is processed using a deconvolution operation with $(2\times2)$ kernel to incorporate the necessary dimensional up-scaling for further processing in the following unit cell.

%\subsection{Proposed Transition Unit}

%\subsubsection{Down transition}
%\subsubsection{Up Transition}

\subsection{Proposed Multi-Scale Fusion (MSF) Module with Pyramid Fusion scheme}
During sequential encoding-decoding operations, a semantic gap is generated between a similar scale of encoded and decoded feature maps. Moreover, in traditional architecture, the gradient has to propagate sequentially that sometimes gives rise to vanishing gradient problems for deeper encoder/decoder module particularly. As multiple stages of encoding and decoding operations are integrated into the CovSegNet, this problem is supposed to be more prominent if all the encoder and decoder modules are sequentially connected. To overcome these limitations, a multi-scale fusion module is proposed that develops parallel interconnection among different scales of feature maps of the encoder/decoder modules utilizing a pyramid fusion scheme.

% \usepackage{multirow}
\begin{table*}[t]
\centering
\caption{\textbf{Ablation Study of the Effect of Different Modules in the Performance (Mean$\pm$Standard Deviation) of the Proposed CovSegNet2D Architecture}}
\label{t3}
\scalebox{0.8}{
\begin{tabular}{|c|c|c|c|c|c|c|c|c|c|c|}
\hline
\multirow{2}{*}{\textbf{Network}} & \multicolumn{5}{c|}{\textbf{Dataset-1}}                                                          & \multicolumn{5}{c|}{\textbf{Dataset-2}}                                                                               \\ \cline{2-11} 
                                  & \textbf{Sen.(\%)} & \textbf{Spec.(\%)} & \textbf{Dice(\%)} & \textbf{IoU(\%)} & \textbf{p-Value} & \textbf{Sen.(\%)} & \textbf{Spec.(\%)} & \textbf{Dice(\%)} & \textbf{IoU(\%)} & \multicolumn{1}{l|}{\textbf{p-Value}} \\ \hline
\textbf{Baseline (V1)}                 & 82.7$\pm$ 0.49         & 97.4$\pm$ 0.09          & 84.1$\pm$0.29         & 79.8$\pm$0.21        & -                & 71.7$\pm$0.12         & 95.8$\pm$0.18          & 71.9$\pm$0.33         & 65.8$\pm$0.27        & -                                     \\ \hline
\textbf{Baseline+ DT (V2)}             & 83.8$\pm$0.29         & 97.8$\pm$0.12          & 85.8$\pm$0.36         & 81.1$\pm$0.08        & 0.0033           & 73.6$\pm$0.31         & 96.5$\pm$0.15          & 73.4$\pm$0.14         & 67.6$\pm$0.21        & 0.0023                                \\ \hline
\textbf{Baseline+ UT (V3)}             & 83.1$\pm$0.25         & 97.7$\pm$0.08          & 85.4$\pm$0.16         & 80.9$\pm$0.13        & 0.0017           & 73.1$\pm$0.55         & 96.3$\pm$0.18          & 73.1$\pm$0.19         & 67.2$\pm$0.35        & 0.0044                                \\ \hline
\textbf{Baseline+ DT+UT (V4)}          & 84.9$\pm$0.41         & 98.1$\pm$0.11          & 86.7$\pm$0.27         & 82.3$\pm$0.32        & 0.0021           & 74.6$\pm$0.17         & 97.1$\pm$0.12          & 74.8$\pm$0.34         & 69.4$\pm$0.18        & 0.0012                                \\ \hline
\textbf{Baseline+(MSF-w/o PF) (V5)}    & 86.9$\pm$0.15         & 98.3$\pm$0.07          & 87.3$\pm$0.28         & 82.9$\pm$0.26        & 0.0019           & 76.2$\pm$0.27         & 97.9$\pm$0.16          & 77.2$\pm$0.29         & 72.8$\pm$0.24        & 0.0034                                \\ \hline
\textbf{Baseline+ MSF (V6)}            & 88.4$\pm$0.28         & 98.7$\pm$0.08          & 89.2$\pm$0.32         & 84.1$\pm$0.21        & 0.0041           & 78.8$\pm$0.25         & 98.4$\pm$0.11          & 79.5$\pm$0.21         & 74.1$\pm$0.25        & 0.0048                                \\ \hline
\textbf{CovSegNet2D (V7)}              & \textbf{90.8$\pm$0.32}         & \textbf{99.1$\pm$0.13}          & \textbf{91.1$\pm$0.25}         & \textbf{86.9$\pm$0.09}        & \textbf{0.0011}           & \textbf{81.5$\pm$0.22}         & \textbf{98.9$\pm$0.13}          & \textbf{82.7$\pm$0.08}         & \textbf{77.5$\pm$0.14}        & \textbf{0.0009}                                \\ \hline
\end{tabular}
}
\end{table*}

% \usepackage{multirow}
\begin{table*}[t]
\centering
\caption{\textbf{Ablation Study of the Effect of Different Modules in the Performance (Mean$\pm$Standard Deviation) of the Proposed CovSegNet3D Architecture in Dataset-1}}
\label{re1}
\scalebox{0.8}{
\begin{tabular}{|c|c|c|c|c|c|}
\hline
\multirow{2}{*}{\textbf{Network}}     & \multicolumn{5}{c|}{\textbf{Dataset-1}}                                                                             \\ \cline{2-6} 
                                      & \textbf{Sensitivity(\%)} & \textbf{Specificity(\%)} & \textbf{Dice Score(\%)} & \textbf{IoU(\%)} & \textbf{p-Value} \\ \hline
\textbf{Baseline3D (V1$_\mathbf{3D}$)}              & 84.5$\pm$0.21                & 97.9$\pm$0.12                & 85.2$\pm$0.23               & 80.8$\pm$0.32          & -                \\ \hline
\textbf{Baseline3D + DT (V2$_\mathbf{3D}$)}         & 85.7$\pm$0.31                & 98.2$\pm$0.19                & 86.1$\pm$0.25               & 82.3$\pm$0.29        & 0.0011           \\ \hline
\textbf{Baseline3D + UT (V3$_\mathbf{3D}$)}         & 85.2$\pm$0.18                & 98.1$\pm$0.08                & 85.9$\pm$0.18               & 82.0$\pm$0.21        & 0.0008           \\ \hline
\textbf{Baseline3D + DT+UT (V4$_\mathbf{3D}$)}      & 86.7$\pm$0.22                & 98.7$\pm$0.14                & 88.3$\pm$0.28               & 83.5$\pm$0.27        & 0.0017           \\ \hline
\textbf{Baseline3D+(MSF-w/o PF) (V5$_\mathbf{3D}$)} & 87.4$\pm$0.25                & 97.9$\pm$0.11                & 88.2$\pm$0.21               & 83.8$\pm$0.31        & 0.0032           \\ \hline
\textbf{Baseline3D+ MSF (V6$_\mathbf{3D}$)}         & 89.6$\pm$0.19                & 98.4$\pm$0.15                & 89.9$\pm$0.17               & 85.1$\pm$0.19        & 0.0021           \\ \hline
\textbf{CovSegNet3D}            & 91.1$\pm$0.26                & 99.3$\pm$0.09                & 92.3$\pm$0.15               & 87.7$\pm$0.23        & 0.0025           \\ \hline
\end{tabular}
}
\end{table*}

% \usepackage{multirow}
\begin{table*}[t]
\centering
\caption{\textbf{Performance Comparison (Mean$\pm$Standard Deviation) of the Proposed CovsegNet2D Architecture  with Other State-of-the-Art Approaches on 2D-CT slices}}
\label{t4}
\scalebox{0.8}{
\begin{tabular}{|c|c|c|c|c|c|c|c|c|c|c|}
\hline
\multirow{2}{*}{\textbf{Network}} & \multicolumn{5}{c|}{\textbf{Dataset-1}}                                                          & \multicolumn{5}{c|}{\textbf{Dataset-2}}                                                                               \\ \cline{2-11} 
                                  & \textbf{Sen.(\%)} & \textbf{Spec.(\%)} & \textbf{Dice(\%)} & \textbf{IoU(\%)} & \textbf{p-Value} & \textbf{Sen.(\%)} & \textbf{Spec.(\%)} & \textbf{Dice(\%)} & \textbf{IoU(\%)} & \multicolumn{1}{l|}{\textbf{p-Value}} \\ \hline
\textbf{Unet~\cite{unet}}                     & 75.9$\pm$0.34         & 88.9$\pm$0.12          & 82.3$\pm$0.26         & 77.1$\pm$0.18        & -                & 52.9$\pm$0.29         & 86.2$\pm$0.09          & 43.3$\pm$0.34         & 38.8$\pm$0.32        & -                                     \\ \hline
\textbf{Unet++~\cite{unet++}}                  & 78.6$\pm$0.17         & 91.1$\pm$0.18          & 84.1$\pm$0.23         & 78.9$\pm$0.21        & -                & 57.7$\pm$0.32         & 89.2$\pm$0.11          & 52.3$\pm$0.31         & 48.1$\pm$0.37        & -                                     \\ \hline
\textbf{MultiResUnet~\cite{mrunet}}             & 77.2$\pm$0.33         & 90.3$\pm$0.24          & 83.7$\pm$0.28         & 78.4$\pm$0.15        & -                & 56.9$\pm$0.27         & 86.9$\pm$0.15          & 50.8$\pm$0.28         & 45.2$\pm$0.22        & -                                     \\ \hline
\textbf{Attention-Unet-2D~\cite{att}}           & 81.1$\pm$0.29         & 92.2$\pm$0.11          & 85.1$\pm$0.14         & 79.6$\pm$0.28        & -                & 60.8$\pm$0.25         & 88.4$\pm$0.12          & 57.7$\pm$0.36         & 51.9$\pm$0.26        & -                                     \\ \hline
\textbf{CPF-Net~\cite{cpf}}                  & 78.9$\pm$0.27         & 91.7$\pm$0.14          & 84.4$\pm$0.25         & 79.3$\pm$0.25        & -                & 62.2$\pm$0.14         & 91.1$\pm$0.14          & 60.4$\pm$0.25         & 56.1$\pm$0.21        & -                                     \\ \hline
\textbf{Semi-Inf-Net~\cite{inf}}             & 82.7$\pm$0.26         & 94.8$\pm$0.21          & 86.9$\pm$0.34         & 81.1$\pm$0.18        & -                & 72.9$\pm$0.44         & 95.8$\pm$0.19          & 74.1$\pm$0.24         & 68.1$\pm$0.32        & -                                     \\ \hline
\textbf{CovSegNet2D(Ours)}              & \textbf{90.8$\pm$0.32}         & \textbf{99.1$\pm$0.13}          & \textbf{91.1$\pm$0.25}         & \textbf{86.9$\pm$0.09}        & \textbf{0.0008}           & \textbf{81.5$\pm$0.22}         & \textbf{98.9$\pm$0.13}          & \textbf{82.7$\pm$0.08}         & \textbf{77.5$\pm$0.14}        & \textbf{0.0013}                                \\ \hline
\end{tabular}
}
\end{table*}

\begin{table}[t]
\centering
\caption{\textbf{Performance Comparison (Mean$\pm$Standard Deviation) of the CovSegNet3D Architecture with Other State-of-the-Art Networks on 3D-CT Volumes of Dataset-1}}
\label{t5}
\scalebox{0.75}{
\begin{tabular}{|c|c|c|c|c|c|}
\hline
\textbf{Network}                                & \textbf{Sen.(\%)} & \textbf{Spec.(\%)} & \textbf{Dice(\%)} & \textbf{IoU(\%)} & \textbf{p-Value} \\ \hline
\textbf{Unet-3D~\cite{unet}}                                & 77.1$\pm$0.22                & 89.8$\pm$0.18                & 84.2$\pm$0.27         & 79.4$\pm$0.24        & -                \\ \hline
\textbf{Unet++-3D~\cite{unet++}}                              & 79.2$\pm$0.17                & 91.7$\pm$0.25                & 85.1$\pm$0.29         & 80.2$\pm$0.26        & -                \\ \hline
\textbf{MultiResUnet-3D~\cite{mrunet}}                        & 78.7$\pm$0.27                & 90.9$\pm$0.16                & 84.5$\pm$0.31         & 78.9$\pm$0.18        & -                \\ \hline
\textbf{Attention-Unet-3D~\cite{att}}                      & 82.5$\pm$0.26                & 93.1$\pm$0.31                & 85.9$\pm$0.24         & 81.4$\pm$0.29        & -                \\ \hline
\textbf{CPF-Net-3D~\cite{cpf}}                             & 80.1$\pm$0.23                & 92.6$\pm$0.23                & 85.2$\pm$0.18         & 80.8$\pm$0.34        & -                \\ \hline
\textbf{VNet-3D~\cite{vnet}}                               & 84.3$\pm$0.29                & 93.9$\pm$0.17                & 85.7$\pm$0.31         & 81.3$\pm$0.19        & -                \\ \hline
\textbf{CovSegNet3D(Ours)}                            & 91.1$\pm$0.26                & 99.3$\pm$0.09                & 92.3$\pm$0.15         & 87.7$\pm$0.23        & 0.0024           \\ \hline
\multicolumn{1}{|l|}{\textbf{CovSegNet-Hybrid(Ours)}} & \textbf{92.6$\pm$0.25}                & \textbf{99.5$\pm$0.07}                & \textbf{94.1$\pm$0.19}         & \textbf{90.2$\pm$0.27}        & \textbf{0.0011}           \\ \hline
\end{tabular}
}
\end{table}

As shown in Fig.~\ref{f6}, each MSF module consists of several MSF-unit cells where each cell considers multi-scale feature maps generated from different levels of preceding encoder/decoder modules and generates feature map for the unit cell of the following encoder/decoder module. Here, similar scale of feature representations generated from different levels of the preceding encoder/decoder modules are concatenated, firstly, to produce $L$ number of multi-scale feature maps. Afterward, all the $L$ scales of feature maps are made spatially equivalent in dimension through pooling and bi-linear upsampling with different windows, and channelwise feature concatenation is carried out to generate the aggregated feature vector. This can be represented as  
\begin{align}
    \nonumber  F_{agg, MSF}^{(i,j)} = P^{(2^{i-1}\times 2^{i-1})} (f_1) \oplus \dots \oplus P^{(2\times 2)} (f_{i-1}) \oplus \\ \oplus  f_i \oplus U^{(2\times 2)} (f_{i+1}) \oplus \dots \oplus U^{(2^{L-i}\times 2^{L-i})}(f_L) \\
    f_{i} = E_{(1, j)} \oplus \dots E_{(i,j)} \oplus D_{(1,j)}\oplus \dots \oplus D_{(i-1,j)}
\end{align}
where $F_{agg, MSF}^{(i,j)}$ is the aggregated feature vector generated in the $i_{th}$ level of $j_{th}$ MSF module, and $f_{i}$ represents the $i_{th}$ concatenated feature map. 

Afterward, the aggregated feature vector is passed through a pyramid fusion scheme to generate the output feature vector that will be fed to the corresponding encoder/decoder unit cell of the following module. Hence, the generated output feature map from each MSF unit cell contains information from all preceding modules and thus, establishes a parallel flow of optimization for efficient gradient propagation.

%The generated encoded and decoded feature maps from eac 

\subsection{Proposed Pyramid Fusion (PF) Module}
The pyramid fusion (PF) module incorporates pyramid fusion scheme into the aggregated feature map of MSF unit cell $(F_{agg, MSF})$ utilizing the combinations of sequential multi-window pooling and upsampling operations (shown in Fig.~\ref{f7}). Firstly, the depth of the aggregated vector, $F_{agg, MSF}$, is reduced through a pointwise convolution (kernel, $1\times1)$ to generate feature vector $f_a$, and thus, $F_{agg, MSF} \mapsto f_a$, where $f_a \in \R^{h \times w \times c}$. 

Afterwards, the generated vector, $f_a$, passes through multiple spatial scaling-vertical scaling-inverse spatial scaling operations in parallel with different scaling factors. 
Spatial scaling operation is carried out utilizing pair of pooling and upsampling operations with different kernel windows, while vertical scaling is employed utilizing convolutional filtering (kernel, $3\times 3$) to reduce the channel depth by one-fourth of the initial depth. Initial reduction followed by expansion of the feature map assists in gathering the more general feature representation, while initial expansion followed by reduction of the feature map gathers the more detailed information from a sparser domain. These operations pave the way to extract the most generalized representations through analyzing from diverse feature domains, which can be represented by  
\begin{align}
P_r :& \R^{h \times w \times c}\rightarrow \R^{h*r \times w*r \times c} \rightarrow \R^{h*r \times w*r \times c/4} \rightarrow \R^{h \times w \times c/4}  \nonumber\\
&\forall r= \{0.25, 0.5, 2, 4\}
\end{align}
where $P_r$ denotes one of the parallel operational paths in the PF module with a spatial scaling factor of $r$.

Afterwards, feature aggregation operation is carried out utilizing different representations generated at multiple paths along with the input representation to generate the aggregated vector $F_{agg,PF}$, where $F_{agg, PF} \in \R^{h \times w \times 2c}$. Finally, a final pointwise convolution (kernel, $1\times1$) is carried out to generate the output feature map $f_{out,PF}$, where $f_{out, PF} \in \R^{h \times w \times c}$.

%$F_{agg, PF} \rightarrow f_{out}$, such that 

% \usepackage{multirow}
\begin{table}[t]
\centering
\caption{\textbf{Effect of Vertical Expansions (Levels) and Horizontal Expansions (Stages) on the Dice Score (Mean$\pm$Standard Deviation) in Dataset-1}}
\label{t6}
\scalebox{0.78}{
\begin{tabular}{|c|c|c|c|c|c|c|}
\hline
\multirow{2}{*}{\textbf{Level}} & \multicolumn{3}{c|}{\textbf{CovSegNet2D}}                & \multicolumn{3}{c|}{\textbf{CovSegNet3D}}                \\ \cline{2-7} 
                                & \textbf{1-stage} & \textbf{2-stage}   & \textbf{3-stage} & \textbf{1-stage} & \textbf{2-stage}   & \textbf{3-stage} \\ \hline
\textbf{2}                      & 49.9$\pm$0.37        & 75.3$\pm$0.13          & 78.12$\pm$0.21       & 57.3$\pm$0.18        & 79.8$\pm$0.18          & 82.1$\pm$0.19        \\ \hline
\textbf{3}                      & 64.8$\pm$0.23        & 85.8$\pm$0.32          & 88.5$\pm$0.15        & 69.3$\pm$0.35        & 89.2$\pm$0.26          & 90.2$\pm$0.25        \\ \hline
\textbf{4}                      & 75.2$\pm$0.32        & 89.6$\pm$0.27          & 90.8$\pm$0.22        & 79.8$\pm$0.29        & \textbf{92.3$\pm$0.15} & 91.8$\pm$0.17        \\ \hline
\textbf{5}                      & 83.5$\pm$0.19        & \textbf{91.1$\pm$0.25} & 89.9$\pm$0.12        & 84.5$\pm$0.43        & 90.2$\pm$0.34          & 89.7$\pm$0.28        \\ \hline
\textbf{6}                      & 86.7$\pm$0.27        & 90.9$\pm$0.21          & 89.1$\pm$0.11        & 89.3$\pm$0.21        & 89.8$\pm$0.41          & 87.9$\pm$0.36        \\ \hline
\end{tabular}
}
\end{table}

\subsection{Structure of the Joint Optimizer}
The decoded feature maps generated from the top of decoder modules are considered for final reconstruction through a joint optimization process. This process is schematically shown in Fig.~\ref{f8}. Initially, an aggregated feature vector $F_{agg, \mathcal{J}}$, is created considering all the output feature maps from different decoder modules which can be given by
\begin{align}
F_{agg, \mathcal{J}} = D_{1,1} \oplus D_{1,2} \oplus \dots \oplus D_{1,S}
\end{align}
where $S$ denotes total number of stages.

Afterward, pyramid fusion scheme is employed on aggregated vector to obtain the more generalized representation utilizing multi-scale decoded representations. Finally,  another convolutional filtering (kernel, $3\times3$) is carried out to generate the final segmentation mask $f_{mask}$, utilizing  binary activation function, and these can be represented as
\begin{equation}
    f_{mask} = \sigma(Conv(PF(F_{agg, \mathcal{J}}))
\end{equation}
where $\sigma(.)$ denotes the non-linear activation.

\subsection{Loss Function}
%Though dice loss function is widely used in medical image segmentation, it suffers for its equal weighting to false positives and false negatives in small region-of-interest with unbalanced classes.  
Tversky Index is introduced in ~\cite{tversky} for better generalization of the the dice index by balancing out false positives and false negatives, which is given by
\begin{equation}
    TI = \frac{\sum_{i=1}^P p_{1i}g_{1i} + \epsilon}{\sum_{i=1}^P p_{1i}g_{1i} + \alpha \sum_{i=1}^P p_{0i}g_{1i} + \beta \sum_{i=1}^P p_{1i}g_{0i} + \epsilon}
\end{equation}
where $g_{0i}$, $p_{0i}$ indicate the ground truth and prediction probability of pixel $i$ being in a normal region, while $g_{1i}$, $p_{1i}$ indicate the ground truth and prediction probability of pixel $i$ being in an abnormal region, $P$ is the total number of pixels on a certain image, $\alpha$, $\beta$ are used to shift emphasize for balancing class imbalance such that $\alpha + \beta =1$, and $\epsilon(10^{-8})$ is used to avoid division-by-zero as safety factor.

To put more emphasis on hard training examples, a Focal Tversky loss function is introduced in~\cite{focal}  utilizing the Tversky Index, which is given by
\begin{equation}
    \mathcal{L} = \sum_c (1 - TI_c)^{\frac{1}{\gamma}}
\end{equation}
where $\gamma$ is used to emphasize the challenging less accurate predictions. Due to the better generalization over a large number of datasets according to~\cite{focal}, $\alpha=0.7, \beta=0.3, \gamma=\frac{4}{3}$ are used for all experimentations in this study.

If $\mathbf{y,y^p}$ denote slice-wise mask ground truth and corresponding probability prediction, respectively, while $\mathbf{Y,Y^p}$ denote volumetric mask ground truth and corresponding probability prediction, respectively, the objective loss functions for separately optimizing CovSegNet2D and CovSegNet3D can be represented as
\begin{align}
    &\mathscr{L}_{2D} = \mathcal{L}(\mathbf{y, y^p});\ \mathbf{y,y^p} \in \R^{h\times w \times c}\\
    &\mathscr{L}_{3D} = \mathcal{L}(\mathbf{Y, Y^p});\ \mathbf{Y,Y^p} \in \R^{h\times w \times s \times c}
\end{align}

\begin{figure}[t]
    \centering
    \includegraphics[scale=0.27]{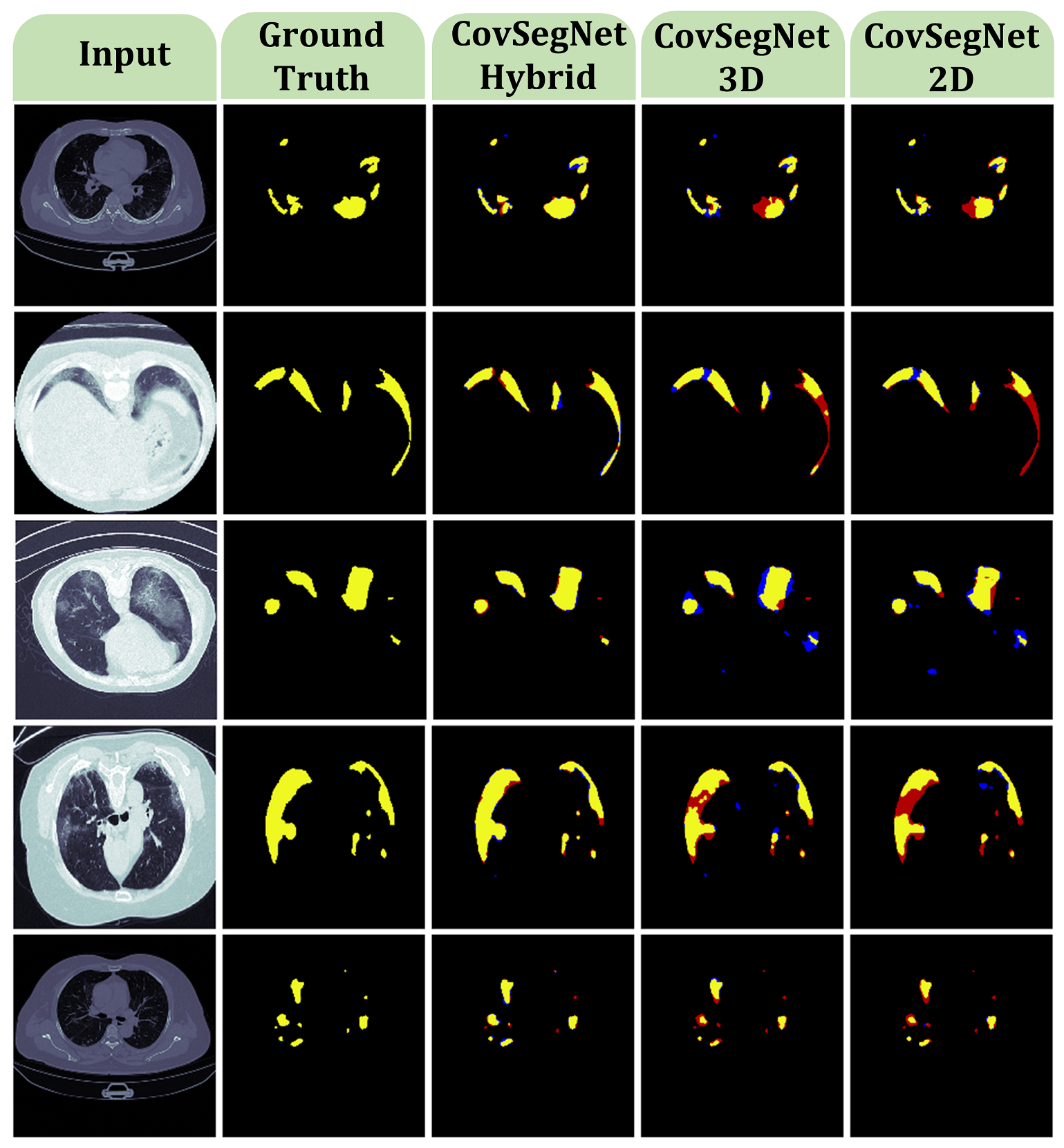}
    \caption{\textbf{Visual representations of the segementation performances obtained using single phase training (CovSegNet2D and CovSegNet3D) and multi-phase training (with hybrid 2D-3D networks) in Dataset-1. Here, `yellow' represents the true positive (TP) regions, `red' represents the false negative (FN) regions, and `blue' represents the false positive (FP) regions.}}
    \label{f10}
\end{figure}

The joint optimization objective function used in phase-2 combining slice-wise and volumetric operations is given by
\begin{equation}
    \mathcal{F} = \lambda (\frac{1}{S}\sum_{i=1}^S \mathscr{L}_{2D}^i) + \mathscr{L}_{3D} 
    \label{joint}
\end{equation}
where $\lambda$ denotes the scaling factor of 2D-loss term, and $s$ denotes total number of 2D-slices per volume. Here, $\lambda = 0.2$ is used for optimization to provide more emphasis on CovSegNet3D in phase-2 as CovSegNet2D is pre-trained in phase-1 and is supposed to be fine-tuned in phase-2.

\section{Results and Discussions}
Experimentations have been carried out on three publicly available datasets to validate the effectiveness of the proposed scheme on numerous segmentation tasks. Performances of CovSegNet2D and CovSegNet3D have been separately studied along with the proposed hybrid scheme of joint optimization combining CovSegNet2D and CovSegNet3D.

\subsection{Dataset Description}
%20 volume
Dataset-1 contains 20 CT volumes with 1800+ slices annotated by expert radiologist panel~\cite{d1}. All the slices have annotations for both lung and infection regions. Each slices are of resolution $(630 \times 630)$ which are resized to $(512\times 512)$. Dataset-2 is the ``COVID-19 CT Segmentation dataset" that contains 110 axial CT images collected by the Italian Society of Medical and Interventional Radiology from 40 different COVID-patients~\cite{d2}. All the images are of resolution $(512\times 512)$. Each slice contains multi-class annotations of infections. Dataset-3 is the ``Semantic Drone Dataset'' where the semantic understanding of urban scenes is mainly focused to increase the safety of drone flight and landing procedures~\cite{d3}. This dataset consists of 400 images with pixel-wise annotation for 20 different classes having resolutions of $6000\times4000$ and all of these images are resized to $(512\times 512)$. Experimentations on Dataset-3 is mainly integrated to investigate the effectiveness of the proposed CovSegNet architecture on other domains with challenging operating conditions. 
%100slices

% \usepackage{multirow}
\begin{table*}[t]
\centering
\caption{\textbf{Comparison of Performances (Mean$\pm$Standard Deviation) on Different Types of Infections (Ground Glass Opacity and Consolidation) in Different CT-slices of Dataset-2}}
\label{t7}
\scalebox{0.9}{
\begin{tabular}{|c|c|c|c|c|c|c|c|c|c|c|}
\hline
\multirow{2}{*}{\textbf{Network}} & \multicolumn{5}{c|}{\textbf{Consolidation}}                                                      & \multicolumn{5}{c|}{\textbf{Ground-Glass Opacity}}                                               \\ \cline{2-11} 
                                  & \textbf{Sen.(\%)} & \textbf{Spec.(\%)} & \textbf{Dice(\%)} & \textbf{IoU(\%)} & \textbf{p-Value} & \textbf{Sen.(\%)} & \textbf{Spec.(\%)} & \textbf{Dice(\%)} & \textbf{IoU(\%)} & \textbf{p-Value} \\ \hline
\textbf{Unet~\cite{unet}}                     & 41.1$\pm$0.26         & 96.2$\pm$0.12          & 40.3$\pm$0.28         & 35.5$\pm$0.28        & -                & 35.1$\pm$0.27         & 98.2$\pm$0.09          & 44.1$\pm$0.27         & 39.8$\pm$0.25        & -                \\ \hline
\textbf{Unet++~\cite{unet++}}                   & 48.8$\pm$0.23         & 97.8$\pm$0.16          & 42.6$\pm$0.26         & 38.2$\pm$0.19        & -                & 41.2$\pm$0.32         & 96.6$\pm$0.14          & 49.9$\pm$0.22         & 45.7$\pm$0.27        & -                \\ \hline
\textbf{MultiResUnet~\cite{mrunet}}             & 46.6$\pm$0.28         & 97.1$\pm$0.14          & 42.1$\pm$0.19         & 37.6$\pm$0.27        & -                & 44.5$\pm$0.28         & 97.3$\pm$0.11          & 47.7$\pm$0.18         & 43.1$\pm$0.28        & -                \\ \hline
\textbf{Attention-Unet-2D~\cite{att}}           & 44.8$\pm$0.19         & 96.8$\pm$0.08          & 44.5$\pm$0.25         & 40.1$\pm$0.33        & -                & 55.3$\pm$0.31         & 95.4$\pm$0.08          & 52.9$\pm$0.17         & 47.6$\pm$0.35        & -                \\ \hline
\textbf{CPF-Net~\cite{cpf}}                  & 49.9$\pm$0.18         & 97.4$\pm$0.15          & 44.1$\pm$0.23         & 39.9$\pm$0.29        & -                & 53.5$\pm$0.22         & 96.9$\pm$0.13          & 56.9$\pm$0.26         & 51.1$\pm$0.34        & -                \\ \hline
\textbf{Semi-Inf-Net~\cite{inf}}             & 50.9$\pm$0.22         & 96.7$\pm$0.11          & 45.8$\pm$0.31         & 41.4$\pm$0.18        & -                & 62.2$\pm$0.34         & 96.1$\pm$0.18          & 62.7$\pm$0.22         & 58.4$\pm$0.23        & -                \\ \hline
\textbf{CovSegNet2D(Ours)}              & \textbf{63.8$\pm$0.17}         & \textbf{98.4$\pm$0.09}          & \textbf{56.8$\pm$0.24}         & \textbf{51.9$\pm$0.25}        & \textbf{0.0017}           & \textbf{73.3$\pm$0.25}         & \textbf{98.9$\pm$0.12}          & \textbf{70.9$\pm$0.31}         & \textbf{66.1$\pm$0.19}        & \textbf{0.0028}           \\ \hline
\end{tabular}
}
\end{table*}

% \usepackage{multirow}
\begin{table}[t]
\centering
\caption{\textbf{Comparison of Performances (Mean$\pm$Standard Deviation) on Multi-Class Semantic Segmentation Task of Dataset-3}}
\label{non}
\scalebox{0.8}{
\begin{tabular}{|c|c|c|c|c|c|}
\hline
\multirow{2}{*}{\textbf{Network}} & \multicolumn{5}{c|}{\textbf{Dataset-3}}                                                          \\ \cline{2-6} 
                                  & \textbf{Sen.(\%)} & \textbf{Spec.(\%)} & \textbf{Dice(\%)} & \textbf{IoU(\%)} & \textbf{p-Value} \\ \hline
\textbf{Unet~\cite{unet}}                     & 56.9$\pm$0.19         & 68.6$\pm$0.23          & 42.2$\pm$0.35         & 37.7$\pm$0.28        & -                \\ \hline
\textbf{Unet++~\cite{unet++}}                   & 57.3$\pm$0.25         & 70.4$\pm$0.31          & 44.8$\pm$0.29         & 40.1$\pm$0.33        & -                \\ \hline
\textbf{Attention-Unet-2D~\cite{att}}           & 58.7$\pm$0.22         & 71.8$\pm$0.29          & 48.5+0.42         & 43.9$\pm$0.25        & -                \\ \hline
\textbf{CPF-Net~\cite{cpf}}                  & 61.5$\pm$0.28         & 73.1$\pm$0.17          & 51.4$\pm$0.38         & 47.7$\pm$0.34        & -                \\ \hline
\textbf{Semi-Inf-Net~\cite{inf}}             & 64.9$\pm$0.31         & 76.3$\pm$0.27          & 50.9$\pm$0.27         & 46.4$\pm$0.26        & -                \\ \hline
\textbf{CovSegNet(ours)}          & \textbf{76.4$\pm$0.18}         & \textbf{87.7$\pm$0.16}          & \textbf{64.6$\pm$0.21}         & \textbf{59.5$\pm$0.29}        & \textbf{6e-5}             \\ \hline
\end{tabular}
}
\end{table}

\subsection{Experimental Setup}
Different hyper-parameters of the network are chosen through experimentation for better performance. Adam optimizer is employed for optimization of the network during the training phase with an initial learning rate of $10^{-5}$. The learning rate is decayed after ever 10 epochs with a decaying rate of 0.99. Intel\textregistered\  Xeon\textregistered \ $D-1653N$ CPU @$2.80$GHz with $12$M Cache and $8$ cores along with $24$ GB RAM is used for experimentation. For hardware acceleration, $2\times$ NVIDIA RTX $2080$ Ti GPU having with $4608$ CUDA cores running $1770$ MHz with $24$ GB GDDR6 memory is deployed. The network is trained for $1000$ epochs on each dataset. Batch size is chosen to be 32 for processing 2D-CT slices, while it is chosen to be 2 for processing 3D-CT volume.

A number of traditional evaluation metrics are used for the evaluation of performance. These are  given by
\begin{align}
    &IoU = \frac{TP}{TP+FP+FN} \\
    &Dice \ Score = \frac{2TP}{2TP+ FP+FN} \\
    &Specificity = \frac{TP}{TP+FP} \\
    &Sensitivity = \frac{TP}{TP+FN} 
%    &F1 \ Score = \frac{2 \times Precision \times Recall}{Precision+ Recall}
\end{align}
where $TP$, $FP$, $FN$ denote true positive, false positive, and false-negative predictions, respectively. A five-fold cross-validation scheme is carried out separately on these databases for evaluation of the proposed scheme. Mean and standard deviations of the evaluation metrics obtained from different test folds are reported. For binary thresholding of the predicted probability mask, a threshold of 0.5 is used in general. The Wilcoxon rank-sum test is used for statistical analysis of the performance improvement obtained from the proposed scheme. The performances of the proposed schemes are statistically analyzed and the statistical significance level is set to $\alpha = 0.01$. The null hypothesis is that no significant improvement of performance is achieved using the proposed scheme over the other existing best performing approaches.

\begin{figure*}[t]
    \centering
    \includegraphics[scale=0.37]{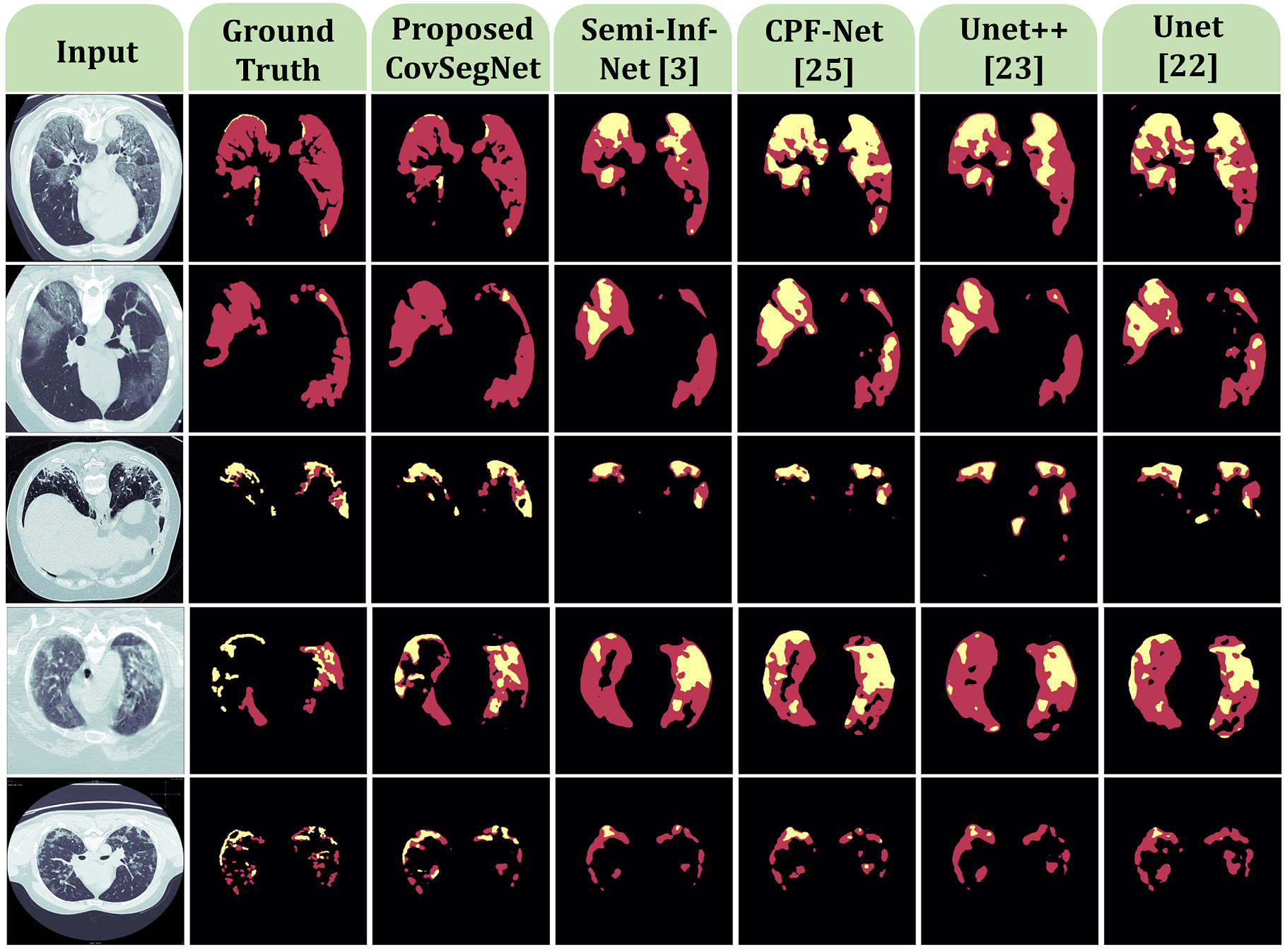}
    \caption{\textbf{Visual representations of the segmented multi-class lesions of the CT images from Database-2 obtained using different state-of-the-art networks. Here, `red' represents the `Ground Glass Opacity (GGO)' regions and `yellow' represents the `Consolidation' regions.}}
    \label{f11}
\end{figure*}

\subsection{Ablation study}
To analyze the effectiveness of different modules of the proposed CovSegNet architecture, an ablation study is carried out. The baseline model is defined as the two-stage implementations with encoder and decoder modules only excluding the down-transition (DT) units, up-transition units (UT), and multi-scale fusion modules. The statistical significance test is carried out to validate the improvement of dice-scores over the baseline model.

\subsubsection{Effects of the transition unit}
Instead of proposed down-transition units and up-transition units, traditional max-pooling and upsampling operations are used, respectively, in the baseline model according to the conventions of traditional Unet architecture. Performances with different combinations of transition units are provided in (V2-V4) of Table~\ref{t3} for 2D analysis. The inclusion of down-transition unit (V2) in encoder modules provides 1.7\% improvement and 1.5\% improvement of dice scores in Database-1 and 2, respectively, over the baseline. Moreover, the inclusion of up-transition unit (V3) in decoder modules provides 1.3\% and 1.2\% improvements of dice scores, while the inclusion of both of the transition units (V4) provide 2.6\% and 2.9\% improvements of dice scores in Database-1 and 2, respectively. Hence, both of the up-transition units and down-transition units are contributing considerable improvements over the baseline performance. Similar improvements can be noticeable for 3D variants of the transition units also (from $V2_{3D}$ to $V4_{3D}$) that are summarized in Table~\ref{re1}. All the improvements are found to be statistically significant ($p<0.01$).  

\begin{table}[t]
\centering
\caption{\textbf{Effect of Different Loss Functions on the Performance (Dice Score(\%)) of CovSegNet on Dataset-1}}
\label{loss}
\scalebox{0.8}{
\begin{tabular}{|c|c|c|c|}
\hline
\textbf{Loss Function}       & \textbf{CovSegNet2D} & \textbf{CovSegNet3D} & \textbf{CovSegNet-Hybrid} \\ \hline
\textbf{Dice Loss}           & 90.2$\pm$0.13            & 91.1$\pm$0.21            & 93.3$\pm$0.09                 \\ \hline
\textbf{Dice Loss+ BCE loss} & 90.4$\pm$0.11            & 91.5$\pm$0.17            & 93.6$\pm$0.15                 \\ \hline
\textbf{Focal Tversky loss}  & 91.1$\pm$0.25            & 92.3$\pm$0.15            & 94.1$\pm$0.19                 \\ \hline
\end{tabular}}
\end{table}

\subsubsection{Effects of the multi-scale fusion (MSF) module}
The MSF modules are proposed in place of the traditional direct skip connection scheme of Unet architecture to reduce the semantic gaps between subsequent encoder and decoder modules. In the baseline model, direct skip connections are used between succeeding modules instead of the MSF module. In Table~\ref{t3}, the change of performance with the inclusion of the MSF module in the 2D-baseline model is provided in V6. It should be noticed that 5.1\% improvement of dice-score, 4.3\% improvement of IoU score have been achieved in Database-1, while 7.6\% improvement of dice-score, 8.3\% improvement of IoU score have been achieved in Database-2. Similar performance improvements can be noticed for the incorporation of MSF module in the 3D-baseline model ($V6_{3D}$ in Table~\ref{re1}). These improvements are found to be statistically significant ($p<0.01$).

\subsubsection{Effects of the pyramid fusion scheme in MSF module}
Pyramid fusion (PF) modules are integrated into the MSF modules to operate on the aggregated multi-scale feature vector in the MSF module. Instead of the PF module, a point-wise convolution with $(1\times1)$ kernel can be performed to reduce and transform the aggregated vector into the output vector. The performance of the 2D-baseline model including this simplified version of the MSF module is reported in V5 of Table~\ref{t3}. It is to be noted that 2.3\% improvement of dice score is achieved in Database-1 and 3.4\% improvement is achieved in Database-2 over the baseline model using these simplified MSF modules, and these improvements are statistically significant ($p<0.01$). However, 3.2\% and 5.3\% reduction of dice scores can be noticed in Database-1 and 2, respectively, from the baseline model with original MSF modules (V6) incorporating PF scheme. Similarly, considerable improvement is also achieved for the incorporation of 3D-pyramid fusion scheme in the 3D variants of MSF module which can be noticed from $V5_{3D}$ and $V6_{3D}$ in Table~\ref{re1}. It justifies the effectiveness of the pyramid fusion scheme in the MSF module.

\subsubsection{Effects of vertical and horizontal scaling}
The proposed CovSegNet architecture is designed in a modular way with the opportunity for both vertical and horizontal expansions for integrating more number of levels and stages, respectively. In Table~\ref{t6}, the performances of the CovSegNet architecture with different numbers of levels and stages are provided. It should be noticed that the optimum dice score of 91.1\% is obtained for CovSegNet2D with 5-levels and 2-stages. The best performance on single stage implementation is found to be 86.7\%, which is 4.4\% lower than the best of the 2-stage implementation. Similar analyses have been carried out on CovSegNet3D using volumetric data where the highest dice score of 92.3\% is achieved with 3-levels and 2-stages implementation.
Moreover, when more stages are included, comparably higher performances are obtained in a lower number of levels, e.g. best dice score of 90.8\% in the 3-stage setup of CovsegNet2D has been achieved with 4-levels.
With the horizontal expansion, the model gathers more amount of contextual information in a lower number of stages that result in higher performances. However, more expansion in both directions starts to increase the complexity that causes a decrease in performance due to overfitting issues.

\subsubsection{Effects of the hybrid 2D-3D joint optimization scheme with two-phase training}
The proposed 2-phase training scheme exploits the advantages of both the slice-based optimization and volumetric optimization. Quantitative performances obtained using CovSegNet2D, CovSegNet3D, and the hybrid scheme are provided in Table~\ref{t4} and \ref{t5}. Slice based processing provides the advantages of employing deeper networks for lighter 2D-convolutions, while loses the contextual information from $z$-axis. On the other hand, volumetric analysis increases the computational burden for optimization for 3D-kernels processing while providing more contextual information. The best variant of CovSegNet3D provides 1.2\% higher dice score, and 0.8\% higher IoU score over the best variant of CovSegNet2D. Thus, the performances of the proposed CovSegNet architectures are quite comparable in both 2D and 3D processing with minor variations. It should be noted that by combining the advantages of both these schemes in the proposed multi-phase training approach, 3\% and 1.8\% higher dice scores are achieved compared to the best performing CovSegNet2D and CovSegNet3D architectures, respectively.  Moreover, to reduce the computational burden of 3D-data processing in the hybrid scheme,  only 2-level, dual-stage implementation of the CovSegNet3D is employed accompanied by the 4-level, dual-stage implementation of the CovSegNet2D that provides the optimal performance with minimal complexity. This improvement signifies the effectiveness of the hybrid networking scheme in multi-phase training ($p<0.01$). Moreover, qualitative analysis of the performances of the individual networks and hybrid networks are presented in Fig.~\ref{f10} with different levels of infection. It should be noticed that both of the false positive and false negative regions are reduced in the segmented mask for the hybrid scheme compared to the individual networks.

\subsubsection{Effects of the loss functions}
In Table~\ref{loss}, effect of different loss functions are summarized on the performance of the CovSegNet. For optimizing the hybrid network, joint optimization objective function (Eqn.~\ref{joint}) is defined incorporating losses of the CovSegNet2D and CovSegNet3D networks. It should be noticed that focal Tversky loss function provides 0.9\% improvement of dice score over traditional dice loss function and 0.7\% improvement over the the aggregated dice loss and binary cross entropy loss function. Similar improvement is also achieved for CovSegNet3D and CovSegNet-hybrid network. However, the proposed CovSegNet architecture mostly provides stable performance over different traditional loss functions, though the optimum performance is achieved with the focal Tversky loss for higher emphasis on the hard training examples.

% \usepackage{multirow}
\begin{table*}[t]
\centering
\caption{\textbf{Computational Efficiency Analysis of Numerous Architectures along with the Performances Obtained on Dataset-1}}
\label{c}
\scalebox{0.73}{
\begin{tabular}{|c|c|c|c|c|c|c|c|c|c|c|c|}
\hline
\multicolumn{6}{|c|}{\textbf{2-Dimensional Analysis}}                                                                                                                                                                                                                                                                                                                                                                                   & \multicolumn{6}{c|}{\textbf{3-Dimensional Analysis}}                                                                                                                                                                                                                                                                                                                                                                                                                                                  \\ \hline
\multirow{2}{*}{\textbf{Architecture}} & \multirow{2}{*}{\textbf{Details}} & \multirow{2}{*}{\textbf{\begin{tabular}[c]{@{}c@{}}Total\\ Parameters(M)\end{tabular}}} & \multirow{2}{*}{\textbf{\begin{tabular}[c]{@{}c@{}}GPU\\ Usage(GB)\end{tabular}}} & \multirow{2}{*}{\textbf{\begin{tabular}[c]{@{}c@{}}Inference\\ Time(s)\end{tabular}}} & \multirow{2}{*}{\textbf{\begin{tabular}[c]{@{}c@{}}Mean\\ Dice(\%)\end{tabular}}} & \multirow{2}{*}{\textbf{Architecture}}                              & \multirow{2}{*}{\textbf{Details}}                                  & \multirow{2}{*}{\textbf{\begin{tabular}[c]{@{}c@{}}Total\\ Parameters(M)\end{tabular}}} & \multirow{2}{*}{\textbf{\begin{tabular}[c]{@{}c@{}}GPU\\ Usage(GB)\end{tabular}}} & \multirow{2}{*}{\textbf{\begin{tabular}[c]{@{}c@{}}Inference\\ Time(s)\end{tabular}}} & \multirow{2}{*}{\textbf{\begin{tabular}[c]{@{}c@{}}Mean\\ Dice(\%)\end{tabular}}} \\
                                       &                                   &                                                                                         &                                                                                      &                                                                                       &                                                                                   &                                                                     &                                                                    &                                                                                         &                                                                                      &                                                                                       &                                                                                   \\ \hline
\textbf{Unet~\cite{unet}}                          & -                                 & 31.0                                                                                    & 2.1                                                                                & 0.10                                                                                  & 82.3                                                                              & \textbf{Unet3D~\cite{unet}}                                                     & -                                                                  & 90.3                                                                                    & 13.2                                                                                & 1.22                                                                                  & 84.2                                                                              \\ \hline
\textbf{Semi-Inf-Net~\cite{inf}}                  & -                                 & 33.3                                                                                    & 6.8                                                                                & 0.18                                                                                  & 86.9                                                                              & \textbf{Vnet3D~\cite{vnet}}                                                     & -                                                                  & 45.1                                                                                    & 15.1                                                                                & 1.16                                                                                  & 85.7                                                                              \\ \hline
\textbf{Unet++~\cite{unet++}}                        & -                                 & 27.0                                                                                    & 6.5                                                                                & 0.17                                                                                  & 84.1                                                                              & \textbf{MultiResUnet3D~\cite{mrunet}}                                             & -                                                                  & 18.1                                                                                    & 12.9                                                                                 & 1.15                                                                                  & 84.5                                                                              \\ \hline
\textbf{CPF-Net~\cite{cpf}}                       & -                                 & 32.4                                                                                    &  2.3                                                                               & 0.12                                                                                  & 84.4                                                                              & \textbf{Attention Unet3D~\cite{att}}                                             & -                                                                  & 103.5                                                                                   & 20.9                                                                                & 1.13                                                                                  & 85.9                                                                              \\ \hline 
\textbf{CovSegNet2D-v1 (ours)}                 & L-2, S-2                          & 0.37                                                                                    & 1.1                                                                                 & 0.05                                                                                  & 75.3                                                                              & \textbf{CovSegNet3D-v1 (ours)}                                              & L-2, S-2                                                           & 1.1                                                                                     & 7.0                                                                                 & 1.02                                                                                  & 79.8                                                                              \\ \hline
\textbf{CovSegNet2D-v2 (ours)}                 & L-3, S-2                          & 1.60                                                                                    & 1.8                                                                                 & 0.07                                                                                  & 85.8                                                                              & \textbf{CovSegNet3D-v2 (ours)}                                              & L-3, S-2                                                           & 4.6                                                                                     & 13.7                                                                                 & 1.21                                                                                  & 89.2                                                                              \\ \hline
\textbf{CovSegNet2D-v3 (ours)}                 & L-4, S-2                          & 6.70                                                                                    & 3.3                                                                                & 0.11                                                                                  & 89.6                                                                              & \textbf{CovSegNet3D-v3 (ours)}                                              & L-4, S-2                                                           & 19.0                                                                                    & 22.2                                                                                 & 1.85                                                                                  & 92.3                                                                              \\ \hline
\textbf{CovSegNet2D-v4 (ours)}                 & L-5, S-2                          & 27.0                                                                                    & 7.0                                                                                & 0.20                                                                                  & 91.1                                                                              & \textbf{\begin{tabular}[c]{@{}c@{}}CovSegNet\\ Hybrid (ours)\end{tabular}} & \begin{tabular}[c]{@{}c@{}}2D(L-4, S-2)\\ 3D(L-2,S-2)\end{tabular} & 7.8                                                                                     & 10.5                                                                                & 1.14                                                                                  & 94.1                                                                              \\ \hline
\end{tabular}
}
\end{table*}

\begin{figure}[t]
    \centering
    \includegraphics[scale=0.28]{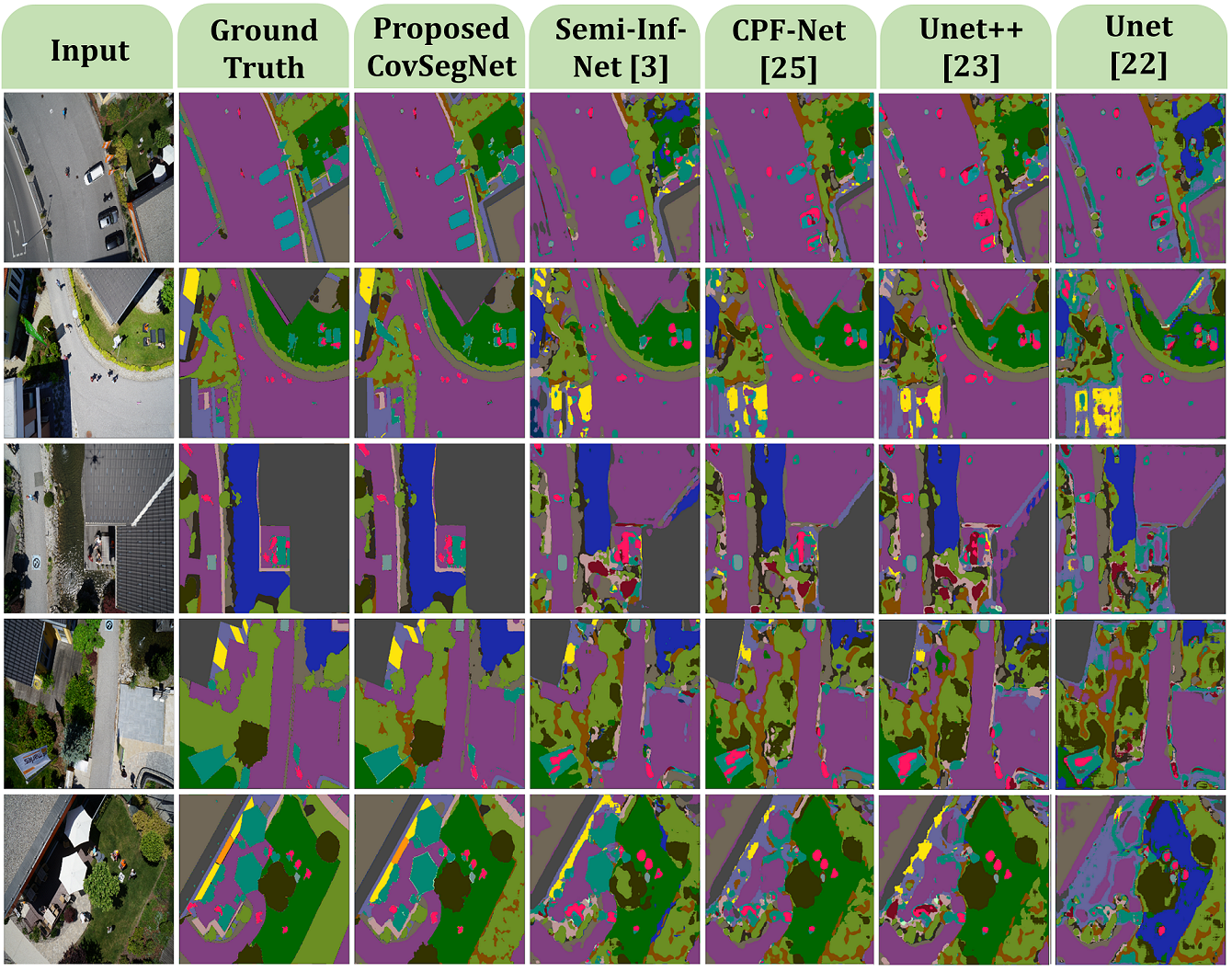}
    \caption{\textbf{Visual representations of the semantic segmentation of drone images from Database-3 obtained using different state-of-the-art networks.}}
    \label{f12}
\end{figure}

\subsection{Comparison with Other Existing Approaches}
To compare the performances of the proposed CovSegNet architecture, several state-of-the-art networks are considered. To compare on a fair platform, most of these networks are implemented using their open-source implementation, and similar train-test folds are used for performance evaluation. Infection segmentation performances using slice-based 2D-operations and volumetric 3D-operations are summarized in Table~\ref{t4} and \ref{t5}, respectively. CovSegNet2D provides a 4.2\% higher dice score in Database-1, and an 8.6\% improvement in dice score in Database-2 compared to the second-highest score (Semi-Inf-Net). Hence, consistent improvements in performances have been achieved in 2D-slice based analysis using CovSegNet2D. Moreover, in the volumetric analysis approach, CovSegNet3D provides an 8.4\% higher dice score and 9.4\% higher IoU score compared to the next-best performing model (VNet). Thus, the 3D-variant of CovSegNet provides consistent improvements over other 3D-counterparts of existing networks.  It should be noticed that the proposed hybrid scheme combining CovSegNet2D and CovSegNet3D provides the most optimum performance with a dice score of 94.1\% and IoU score of 90.2\%. Some of the qualitative visualizations of performances obtained in different challenging conditions are shown in Fig.~\ref{f9}. For having the volumetric information of the Database-1, the proposed hybrid scheme is employed here, while only 2D-slice based analysis is carried out in Database-2 using CovSegNet2D. It should be noted that the proposed scheme performs consistently better compared to other networks in segmenting most of the challenging diffused, blurred, and varying shaped edges of COVID lesions. Moreover, quantitative performances on challenging multi-class lesion segmentation, including separate ground-glass opacity (GGO) and consolidation regions, are summarized in Table~\ref{t7}, where 8.2\% improvement in dice score is obtained in GGO segmentation and 11\% improvement in consolidation segmentation using CovSegNet architecture over the other best-performing approaches.  Additionally, from the visual analysis of the performances shown in Fig.~\ref{f11}, it can be easily noted that the proposed network considerably reduces the false predictions even in these challenging conditions compared to other state-of-the-art approaches.

Furthermore, quantitative results obtained from non-clinical Database-3 are summarized in Table~\ref{non} which shows the significant performance improvement with $22.4\%$ improvement in dice score, and $21.8\%$ improvement in mean IoU compared to the Unet architecture. Weighted mean performances over all 20 classes are taken for better estimation. In Fig.~\ref{f12}, visual representations of some of the sample images are shown for different networks in Database-3, which more conspicuously signifies the better performance of the proposed architecture. Since Database-3 is very complicated with a huge number of classes, the performance differences between the proposed CovSegNet and other existing networks are more prominent as this dataset demands effective exploitation of minute, complex, and scattered features of diversified classes.  

%{Comparison with other existing approaches on Ground glass opacity and consolidation Segmentation}

\subsection{Computational Efficiency Analysis of Numerous Approaches}
The proposed CovSegNet architecture ensures the proper optimization of all the network parameters through improved parallelization that enhances efficient gradient propagation in the whole network resulting in the effective exploitation of the contextual information with consistently good performance. 
However, this improved parallelism also poses some computational burden for the effective exploitation of the network parameters.
%Though this network considerably reduces the network parameters and memory usage, computational complexity seems to rise for efficient exploitation of the available parameters through improved parallelization. 
In Table~\ref{c}, the computational efficiency of different networks are summarized, where performances of different variants of CovSegNet is summarized based on the number of levels ($L$) and stages ($S$). For analyzing with 2D data, it is noticeable that the number of parameters of CovSegNet2D are considerably lower compared to other networks while providing a large improvement of performance. For example, CovSegNet-v2 provides a 94.8\% reduction in parameter counts of Unet architecture, while providing 1.43x higher inference speed with a 3.5\% higher dice score. With increasing levels, more precise estimation is achievable in the cost of speed and memory consumption. Moreover, GPU-memory usage for training with batch size-1 are summarized for different networks. A similar observation can be carried out for 3D analysis with CovSegNet3D. It should be noticed that CovSegNet-Hybrid provides the best achievable dice score (94.1\%) while consisting of 0.09x parameters of Unet3D 
with 0.08$s$ reduction of inference time. This significant reduction in parameter counts with the obtained highest performance is mainly achieved by the joint integration of the efficient 2D processing with effective inter-slice contextual information exploration using a lighter 3D variant of CovSegNet. Therefore, this hybrid scheme provides considerable advantages over other existing 3D variants in terms of parameters, and dice scores with comparable processing speed.

\subsection{Discussions, Limitations, and Future Studies}
In summary, numerous architectural renovations assist in achieving state-of-the-art performance on COVID lesion segmentation. The horizontal and vertical expansion mechanisms provide the opportunity to incorporate more detailed features as well as more generalized features, which improved the feature quality considerably that is particularly effective in distinguishing multi-class, scattered COVID lesions with widely varied shapes. Moreover, the improved gradient flow throughout the network, achieved with the introduction of multi-scale fusion module and scale transition modules, have greatly reduced the contextual information loss in the generalization process and have also ensured the best optimization of all network parameters that particularly contribute to recover and distinguish the blurry, diffused edges of COVID lesions as well as the very minute instances of abnormalities. Furthermore, the integration of a hybrid 2D-3D networking scheme exploits both the intra-slice and inter-slice contextual information without increasing computational burden that results in more precise, finer segmentation performance mostly in challenging conditions.

Although consistent performances have been achieved in both the datasets for COVID lesion segmentation, this study should be carried on larger datasets consisting of wide variations of subjects. However, in the current conditions of the pandemic, it is difficult to gather a considerably higher amount of data. The proposed study will be extended with the incorporation of diversified datasets including patient-based study considering age, sex, health conditions, and geographical locations of the patients. Due to the novel characteristics of the COVID infections, it is difficult to predict the risk and vulnerability among diverse subjects that can be effective for reducing the spread and better prevention. An in-depth, closer, patient-specific study should be carried out for better understandings of the nature of the infection. Moreover, generative adversarial network-based optimization can be carried out to generate more amount of realistic, synthetic data to overcome the limitations of available data. Additionally, this scheme is supposed to be extended for incorporating automated segmentation-classification joint optimization along with the severity prediction scheme of COVID infections. 

\section{Conclusion}
In this study, an automated scheme is proposed with an efficient neural network architecture (CovSegNet) for very precise lung lesion segmentation of COVID CT scans that provides outstanding performances with 8.4\% average improvement of dice score over two datasets.
The introduced scale transition operations are found to be very effective for replenishing contextual information loss through repeated integration of generated multi-scale features in both upscaling and downscaling operations. It is found that horizontal expansion mechanism with multi-stage encoder-decoder modules assists in further improvements for gathering more multi-scale contextual information when coupled with the traditional vertical expansion mechanism. Moreover, the multi-scale fusion module with a pyramid fusion scheme not only substantially reduced the semantic gaps between subsequent encoder-decoder modules but also introduced parallel inter-linking among multi-scale features that greatly mitigates the vanishing gradient issues for better optimization. Furthermore, the two-phase optimization scheme with hybrid 2D-3D processing provides considerable improvement over traditional single domain approaches for introducing more contextual information to gather finer details. It is shown that the proposed scheme is capable of segmenting infected regions along with multi-class COVID-19 lesions with unprecedented precision even in challenging conditions with blurred, diffused, and scattered edges. Moreover, it is found that the proposed network is not only effective in COVID lesion segmentation but also provides state-of-the-art performance on a non-clinical, challenging, multi-class semantic segmentation task that proves the wide applicability of the proposed scheme.
Therefore, the proposed scheme can be easily optimized on numerous applications that can be an effective alternative to other state-of-the-art approaches.

%A two-phase training scheme is introduced for utilizing the advantages of efficient 2D-convolutions with 3D contextual information. A modular architecture is proposed, namely CovSegNet, that is found to be consistently better performing compared to other state-of-the-art approaches in both 2D and 3D domains. Moreover, it is shown that the proposed hybrid scheme provides considerable improvements over single domain approaches. Both horizontal and vertical expansion scheme is incorporated in CovSegNet, where it is found that more number of stages provides better performance than considering single-stage implementations. Parallel optimization of multi-scale features utilizing the proposed MSF module, pyramid fusion scheme, and transition units provides considerable improvements over the baseline architecture. This scheme of multi-phase optimization along with the efficient CovSegNet architecture can be easily optimized on numerous medical imaging applications that can be an effective alternative to other state-of-the-art approaches.    

\bibliographystyle{IEEEtran}
\bibliography{ref}

\end{document}